%% file: manuscript.tex
\documentclass[english,prl,aps,twocolumn,10pt]{revtex4-2}

\usepackage{amsmath}
\usepackage{amssymb}
\usepackage{wasysym}
\usepackage{graphicx}
\usepackage{babel}
\usepackage{multirow}
\usepackage{array}
\usepackage{verbatim}
\usepackage[colorlinks=true, pdfstartview=FitV, linkcolor=blue, citecolor=blue, urlcolor=blue]{hyperref} 
\usepackage{xcolor}
\usepackage{bm}
\usepackage{ulem} 

\def\ie       {{\it i.e.}}

\newcommand{\ee}[1]{\cdot10^{#1}}
\newcommand{\mr}[1]{\mathrm{#1}}
\newcommand{\unit}[1]{\,\mathrm{#1}}
\newcommand{\um}{\,\mu{\rm m}}

\newcommand{\uA}{\,\mu{\rm A}}
\newcommand{\kT}{k_{\rm B}T}

\newcommand{\degree}{^\circ}
\newcommand{\Ohm}{\Omega}

\newcommand{\Bext}{B_\mr{ext}}
\newcommand{\BNVft}{\hat{B}_\mr{NV}}

\newcommand{\Bzft}{\hat{B}_z}
\newcommand{\Bz}{B_z}
\newcommand{\Dv}{D_\nu}
\newcommand{\EF}{E_\mr{F}}
\newcommand{\Io}{I_0}
\newcommand{\Isig}{I_\mr{sig}}
\newcommand{\Ileak}{I_\mr{leak}}
\newcommand{\Jm}{J}
\newcommand{\Jx}{J_x}
\newcommand{\Jy}{J_y}
\newcommand{\Jxft}{\hat{J}_x}
\newcommand{\Jyft}{\hat{J}_y}

\newcommand{\MR}{\textit{MR}}
\newcommand{\nne}{n_\mr{e}}
\newcommand{\nnh}{n_\mr{h}}
\newcommand{\Rc}{R_\mr{c}}
\newcommand{\Vbg}{V_\mr{BG}}
\newcommand{\Vsd}{V_\mr{SD}}

\newcommand{\thetaH}{\theta_\mr{H}}

\begin{document}
	
\title{Direct imaging of magnetotransport at graphene-metal interfaces \\ with a single-spin quantum sensor}

\author{C.~Ding$^{1}$, M.~L.~Palm$^{1}$, K.~Kohli$^{1}$, T.~Taniguchi$^2$, K.~Watanabe$^3$, C.~L.~Degen$^{1,4}$}
\affiliation{$^1$Department of Physics, ETH Z\"urich, Otto Stern Weg 1, 8093 Z\"urich, Switzerland.}
\affiliation{$^2$Research Center for Materials Nanoarchitectonics, National Institute for Materials Science, 1-1 Namiki, Tsukuba 305-0044, Japan.}
\affiliation{$^3$Research Center for Electronic and Optical Materials, National Institute for Materials Science, 1-1 Namiki, Tsukuba 305-0044, Japan.}
\affiliation{$^4$Quantum Center, ETH Z\"urich, 8093 Z\"urich, Switzerland.}
\email{degenc@ethz.ch}

\begin{abstract}	
	Magnetotransport underlines many important phenomena in condensed matter physics, such as the Hall effect and magnetoresistance (MR) effect. Thus far, most magnetotransport studies are based on bulk resistance measurements without direct access to microscopic details of the spatial transport pattern. Here, we report nanoscale imaging of magnetotransport using a scanning single-spin quantum magnetometer, which is demonstrated in a graphene-metal hybrid device at room temperature. By visualizing the current flow at elevated magnetic fields ($\sim 0.5\unit{T}$), we directly observe the Lorentz deflection of current near the graphene-metal interface, which is a hallmark of magnetotransport. Combining the local current distribution with global resistance measurements, we reveal that transport properties of the hybrid are governed by a complex interplay of intrinsic MR around the Dirac cone, carrier hydrodynamics, interface resistance, and the nanoscale device geometry. Furthermore, accessing the local transport pattern across the interface enables quantitative mapping of spatial variations in contact resistance, which is commonly present in electronic devices made from two-dimensional materials yet non-trivial to characterize. Our work demonstrates the potential of nanoscale current imaging techniques for studying complex electronic transport phenomena that are difficult to probe by resistance-based measurements.
\end{abstract}
 
\date{\today}
	
\maketitle
		

The Lorentz force law states that a charged particle moving in a magnetic field experiences a force perpendicular to its velocity and the magnetic field vector.  For free particles, the (magnetic) Lorentz force leads to a deflection of the particle's trajectory.  This phenomenon is observed and exploited in many areas of physics, such as particle accelerators, electron microscopes and mass spectrometers.  In solid-state transport devices, on the other hand, the Lorentz force usually leads to charge accumulation at opposite sides of the device, described by the Hall effect~\cite{hall1879}.  In uniform (semi-)conductors, the ensuing electrostatic force exactly compensates the Lorentz force, and therefore no current deflection is present (Fig.~\ref{fig1}a).

The situation becomes more complex for hybrid devices composed of materials with different electrical mobilities and conductivities.  A well-known example is the semiconductor-metal hybrid, where the metal closely approximates an ideal conductor. At zero external magnetic field, the current is pulled into the metal (Fig.~\ref{fig1}b, left) due to its higher conductivity. By contrast, under an out-of-plane magnetic field, charge carriers are deflected around the metal (Fig.~\ref{fig1}b, right) due to the uncompensated Lorentz force. As the magnetic field strength increases, more current is forced into the semiconductor, resulting in an increase of the device resistance. This phenomenon is known as extraordinary magnetoresistance (EMR).  The EMR is a geometrical MR effect and relies on magnetic-field-induced current deflection at semiconductor-metal interfaces~\cite{solin2000, pomar2024amt}. Despite being fundamental to the magnetotransport properties of such hybrid devices, the current flow pattern is usually not known beyond numerical simulations like the ones shown in Fig.~\ref{fig1}b, and the current deflection has not yet been directly observed.


In this work, we demonstrate direct imaging of magnetotransport in a graphene-metal hybrid using a scanning nitrogen-vacancy (NV) magnetometer. By bringing the nanoscale resolution and ac quantum sensing~\cite{ku2020,vool2021, palm2022, palm2024} to above $0.5\unit{T}$, which to our knowledge is the highest reported bias magnetic field for scanning NV microscopy, we manage to observe key spatial features of magnetotransport, including current deflection due to the Lorentz force and reorganization of currents through the hybrid structure. 
Furthermore, by correlating local transport patterns with global resistance characterization, we find that the magnetotransport physics in the graphene-metal hybrid is surprisingly rich and arises from a combination of the nanoscale device geometry, intrinsic MR around the Dirac cone, carrier hydrodynamics, and contact resistances between conductors.
Our work underscores the importance of material interfaces in magnetotransport and showcases the potential of nanoscale current imaging techniques to address complex transport questions in hybrid devices.


\subsection*{Hybrid device and experimental setup}

\begin{figure*}[ht]
	\centering
	\includegraphics[width=0.85\textwidth]{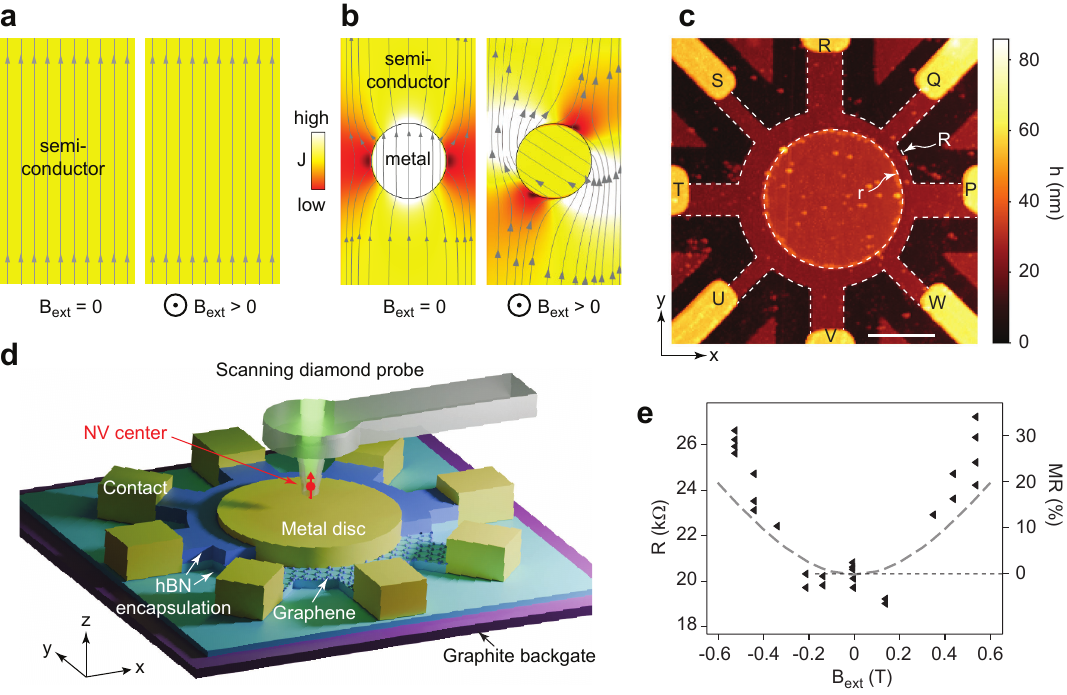}
	\caption{
		\textbf{Concept of the experiment and the graphene-metal hybrid device.}
		{\bf a}, Hall effect in a uniform semiconductor.  No current deflection occurs because the Lorentz force and opposing Hall potential gradient are exactly balanced. $\Bext$ is the out-of-plane external magnetic field.
		{\bf b}, In a semiconductor-metal hybrid, current is attracted to the metal disc at low field (left panel) while being deflected at elevated field (right panel).
		{\bf c},
		Topography of the sample, showing the van-der-Pauw geometry with an inner metal disc (radius $r$), outer graphene ring (radius $R$), and eight contacts labeled \textit{P} to \textit{W}.  Dashed contours mark the edge of the graphene sheet. Scale bar, $1\unit{\um}$.
		{\bf d}, Experimental arrangement for imaging current flow through the device using a scanning NV microscope.
		{\bf e}, Measured two-terminal resistance $R$ (contacts \textit{Q-V}) and computed magnetoresistance $\MR$ (see text), plotted as a function of the out-of-plane magnetic bias field $\Bext$. The dotted line indicates the zero-field resistance and the dashed line represents the simulated resistances~\cite{supplementary}.
	}
	\label{fig1}
\end{figure*}

Our device has a van-der-Pauw disc geometry similar to those used in transport studies of the geometrical MR effect~\cite{solin2000,lu2011,zhou2020apl}.  The device consists of a metallic (Cr/Au) inner disc of radius $r=0.9\unit{\um}$ that is concentric (with a $0.1\unit{\um}$ lateral offset) with the larger monolayer graphene annulus of $R=1.2\unit{\um}$, as shown in~Fig.~\ref{fig1}c.  The graphene sheet is encapsulated in hexagonal boron nitride (hBN) and equipped with a graphite back-gate (Fig.~\ref{fig1}d).  The mobility of the graphene is extracted from resistance measurements to be approximately $\mu_e\approx\mu_h\sim 1.35\pm 0.25\unit{m^2/(Vs)}$~(Methods).  Eight contacts are distributed equally around the disc and allow probing the device at different source-drain configurations in $45^\circ$ steps.

\begin{figure*}[t]
	\includegraphics[width=0.67\textwidth]{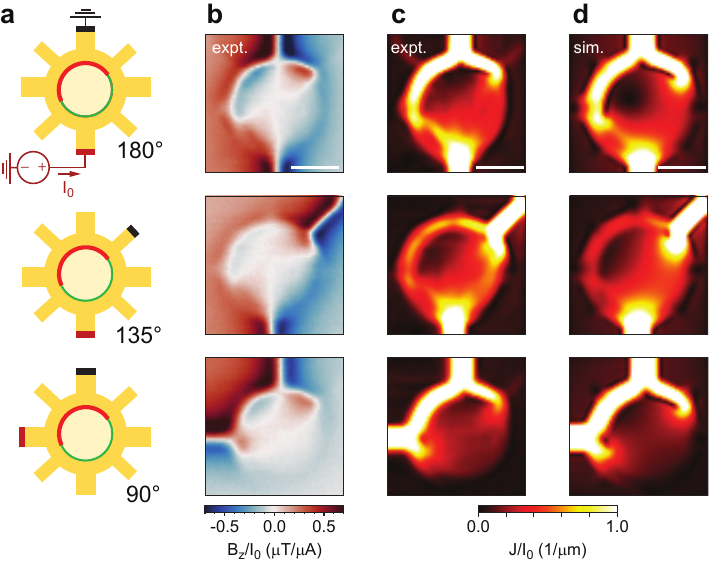}
	\caption{
		\textbf{Imaging of current flow for various source-drain configurations at zero bias field around charge neutrality  ($\Vbg=0$).}
		{\bf a}, Schematic of source-drain configurations. Thick red (thin green) arcs indicate a high (low) contact resistance between the metal disc and the graphene annulus.  See Fig.~S10 for a quantitative model.
		{\bf b}, Measured magnetic field maps (out-of-plane component $\Bz$) induced by a current (amplitude $\Io$) for $\{180^\circ, 135^\circ,90^\circ\}$ contact configurations. 
		{\bf c}, Current density magnitude ($\Jm=\sqrt{\Jx^2+\Jy^2}$) computed from the measured $\Bz$.
		{\bf d}, Simulated current density maps using the quantitative model of Fig.~S10 and Tables S1 and S2.
		Scale bars, $1\unit{\um}$.
	}
	\label{fig2}
\end{figure*}

Fig.~\ref{fig1}d shows a schematic of the imaging setup.  The scanning magnetometer makes use of a diamond probe containing a single nitrogen-vacancy (NV) center near its apex that is scanned in a plane of constant height ($z\sim 100\unit{nm})$ above the graphene layer (Methods). The diamond tip used in this work is fabricated from a (111)-oriented diamond crystal~\cite{rohner2019} such that its anisotropy axis is out-of-plane. This is crucial for maintaining the NV optical contrast and high magnetic field sensitivity in high out-of-plane magnetic bias fields ($|\Bext|\sim 0.5\unit{T}$) required for studying magnetotransport~\cite{epstein2005,tetienne2012}. The bias fields are produced using a stack of permanent magnets located underneath the sample (Methods).

To image the current distribution, we apply an alternating source-drain voltage ($f\sim 30-300\unit{kHz}$) of amplitude $V_\mr{SD}\sim 40-300\unit{mV}$ across the respective contacts to induce a current in the device and detect the associated Oersted field using a spin-echo magnetometry technique (Methods and Ref.~\cite{palm2022}). In addition, we use a differential measurement scheme to reject spurious signals caused by back-gate leakage currents when  the  back-gate voltage is non-zero (Methods and Figs.~S5 and S6).  While imaging, we also record time traces of the source-drain current to monitor its amplitude ($\Io\sim 3-27\unit{\uA}$) and the corresponding two-terminal resistance $R=V_\mr{SD}/\Io$ between the chosen contacts.

The results of a representative $R(\Bext)$ measurement (at $\Vbg=0$) and corresponding unit-less magnetoresistance $\MR = R(\Bext)/R(0)-1$ are shown in Fig.~\ref{fig1}e. We observe typical values of $\MR\sim 30\%$ at $|\Bext|\sim 0.5\unit{T}$.  Note that because we measure a two-terminal resistance, $R(0)$ contains contributions from the source and drain contact resistances ($\Rc\sim 7.5\unit{k\Ohm}$ for contacts \textit{Q-V}~\cite{supplementary}).  With $\Rc$ subtracted, which approximates a four-terminal measurement, $\MR \sim 40\%$ at $0.5\unit{T}$.

\subsection*{Interface resistance and local transport at the graphene-metal boundary}

\begin{figure*}[ht]
	\includegraphics[width=0.99\textwidth]{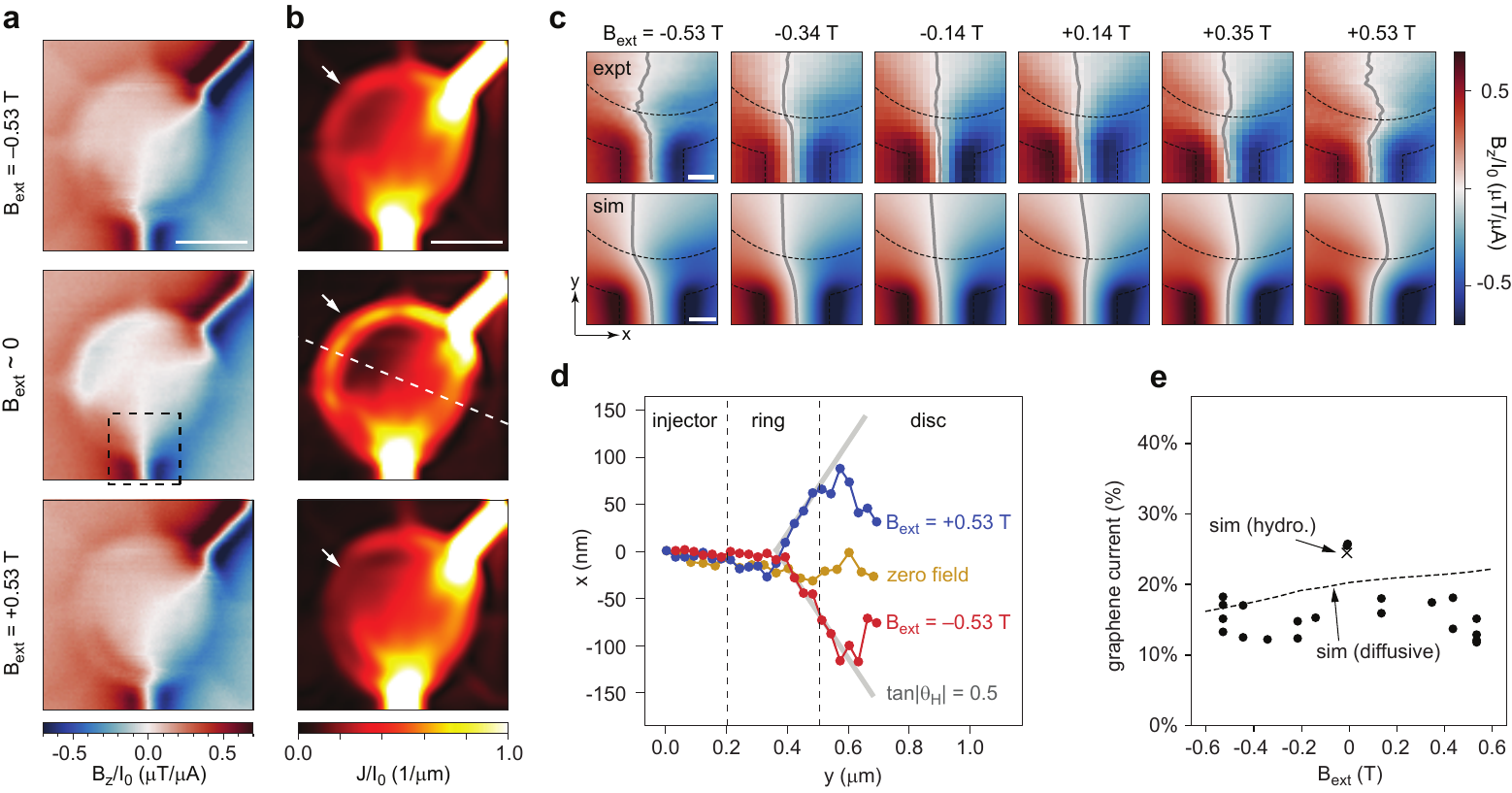}
	\caption{
		\textbf{Lorentz-force-induced deflection and current redistribution under an applied bias field around charge neutrality ($\Vbg=0$)}.
		{\bf a,b}, $\Bz$ maps and corresponding $\Jm$ maps measured at $\Bext = -0.53\unit{T}$ (top), $-0.01\unit{T}$ (center) and $+0.53\unit{T}$ (bottom).  The dataset at $\Bext\sim 0$ is replotted from Fig.~\ref{fig2}. Scale bars, $1~\mr{\um}$.
		{\bf c}, Detail of $\Bz$ (dashed square in panel {\bf a}) revealing Lorentz-induced current deflection.  Dashed contours show the device boundaries.  The gray curves are the center streamlines of the injected current (see text). Scale bars, $200\unit{nm}$.
		{\bf d}, Deflection of the center streamline (color) together with the Hall angle tangent $\tan\thetaH$ (gray), permitting a direct spatial measurement of the carrier mobility $\mu$.
		{\bf e}, Fraction of the current flowing through the graphene ring, obtained by integrating the normal component of the current density along the dashed line in panel {\bf b} (Methods).  Circles are the experimental data and the dashed line corresponds to simulations in the diffusive regime. The cross corresponds to a simulation in the hydrodynamic regime at $\Bext=0$.
	}
	\label{fig3}
\end{figure*}

We begin our spatial transport investigation by imaging the current flow for several contact configurations at zero field ($|\Bext|<0.01\unit{T}$) and at zero back-gate voltage (near the change neutrality point, CNP).  Under these conditions, the MR is absent and the graphene behaves like a semimetal~\cite{castroneto2009} with $\sim 0.5\times$ the conductivity of the metal disc~\cite{supplementary}.
Fig.~\ref{fig2} shows the measured out-of-plane magnetic field maps ($\Bz$) and the associated current density maps ($\Jm$, computed by back-propagation~\cite{roth1989,chang2017,palm2022}) for three source-drain contact configurations probing $180^\circ$, $135^\circ$ and $90^\circ$ sections of the annulus.

We first discuss the $180^\circ$ result.  Here, the current is expected to mainly flow through the metal disc because of the shorter electrical path, in accordance with Fig.~\ref{fig1}b.  Surprisingly, we find the current to branch out into the left and right arms of the graphene annulus, especially in the upper half of the device.  Moreover, there is an unexpected left-right asymmetry.  We attribute both effects to a high contact resistance at the upper left interface.  This picture is corroborated by the current density pattern of the $135^\circ$ and $90^\circ$ contact configurations, which both display negligible current density in the upper left portion of the metal disc.

The data of Fig.~\ref{fig2}b,c and the carrier density dependence (presented below in Fig.~\ref{fig4}) allow us to construct a quantitative conductance model of the device, taking into account the geometry, the bulk conductivities of the metal and graphene layers, and the interface resistances (Fig.~S10 and Tables S1, S2). We optimize this model by matching the experimental current density maps to finite element simulations, reaching excellent agreement (Fig.~\ref{fig2}d). The results confirm the presence of a high contact resistance in the upper left area of the metal disc (marked by thick red arcs in Fig.~\ref{fig2}a) and a low contact resistance in the remaining section~(thin green arcs).

We note that such variations in the contact resistance, while undesired, are difficult to avoid for hybrid device structures~\cite{holz2003,pomar2024arxiv}, especially those involving two-dimensional materials~\cite{wang2013, kwon2022interaction}.  In our device, they are present even if the accompanying topography image (Fig.~\ref{fig1}c) does not reveal any discontinuities at the graphene-metal interface, such as a visible gap.  Although playing a decisive role in transport devices~\cite{moller2003}, contact resistances are notoriously difficult to characterize, especially for lateral contacts~\cite{yang2020pms}.  Our work demonstrates that local imaging techniques are not only suited for investigating device variability, but also allow taking such variability into account for modeling and interpreting the transport behaviors of the device.

\subsection*{Imaging of Lorentz deflection and current redistribution}

Next, we present the main result of our study, the reorganization of current flow when exposing the device to out-of-plane bias fields.  These measurements are performed in the {\it Q-V} contact configuration that mostly avoids the section of high contact resistance.  While the data are taken at $\Vbg=0$, carrier-dependent measurements (Fig.~S8) indicate a slight hole doping.
Figs.~\ref{fig3}a,b show the respective stray field and current density maps.  Interestingly, despite a large MR signature in the corresponding resistance measurement (Fig.~\ref{fig1}e), the current flow patterns are not much affected by the magnetic field. Nevertheless, we observe two features that reveal the complexity of magnetotransport in the graphene-metal hybrid.

The first feature is a magnetic-field-induced current deflection, which is a direct consequence of the Lorentz force and the spatial hallmark of the classical Hall effect.  The current deflection is most obvious near the injection contact (dashed box in Fig.~\ref{fig3}a).
Fig.~\ref{fig3}c plots this region at higher magnification for various bias fields between $\pm 0.53\unit{T}$.  We plot the magnetic stray field map rather than the current map because the $\Bz=0$ contour provides a convenient visual guide of the current's center streamline.  A clear deflection of the center streamline (gray curves) can be noticed. We observe a leftward (rightward) bending of the streamline at negative (positive) bias fields, in agreement with the right hand rule for the Lorentz force law.  Because the deflection is proportional to $\mu\Bext$ for moderate bias fields, we can use the deflection angle (Hall angle $\thetaH$) to extract a value for the carrier mobility that is independent from the transport data (Fig.~\ref{fig3}d).  Specifically, for these measurements, we obtain~\cite{supplementary}
\begin{align}
	\mu &= \left|\frac{\nne+\nnh}{\nne-\nnh}\frac{\tan\thetaH}{\Bext}\right| \sim 1.4\unit{m^2/(Vs)} \ ,
\end{align}
where $\tan|\thetaH|=0.5$, $\Bext=0.53\unit{T}$, $\nne = 0.34\ee{11}\unit{cm^{-2}}$ and $\nnh = 1.74\ee{11}\unit{cm^{-2}}$.  This value for $\mu$ is in good agreement with the mobility inferred from the two-terminal resistance measurement ($\mu\sim 1.35\pm 0.25\unit{m^2/(Vs)}$, see Methods and Fig.~S11).  Our results demonstrate that spatial imaging of magnetotransport provides an alternative means to measure the carrier mobility in hybrid transport devices.

The second feature is a decrease of the current flowing through the graphene ring at elevated fields (marked by arrows in Fig.~\ref{fig3}b), and a corresponding increase of the current within the metal disc.  This runs counter to the generic picture of the EMR effect (Fig.~\ref{fig1}b), where the current is \textit{expelled} from the metal disc at elevated fields.  A quantitative analysis of the relative current contributions, obtained by integrating along the cross-section of the device (dashed line in Fig.~\ref{fig3}b), shows that the fraction of current flowing through the graphene ring drops from $\sim 26\%$ at zero field to $\sim 15\%$ at $|\Bext|>0.1\unit{T}$, and remains approximately constant up to $|\Bext| \sim 0.5\unit{T}$ (Fig.~\ref{fig3}e).
This magnetic field dependence of the graphene current is not correlated with that of $\MR$ (Fig.~\ref{fig1}e) and cannot be explained with the geometrical MR effect.
%
In the following, we argue that this apparent contradiction is resolved by considering two physical mechanisms that are specific to high-quality graphene: (i) the presence of charge hydrodynamics at room temperature~\cite{palm2024} that becomes suppressed in a magnetic field~\cite{berdyugin2019}, and (ii) the strong intrinsic MR effect of charge-neutral graphene~\cite{xin2023} that dominates the $\MR$ vs. $\Bext$ response.

\subsection*{Electron hydrodynamics and single-carrier transport}

\begin{figure*}[ht]
\includegraphics[width=0.85\textwidth]{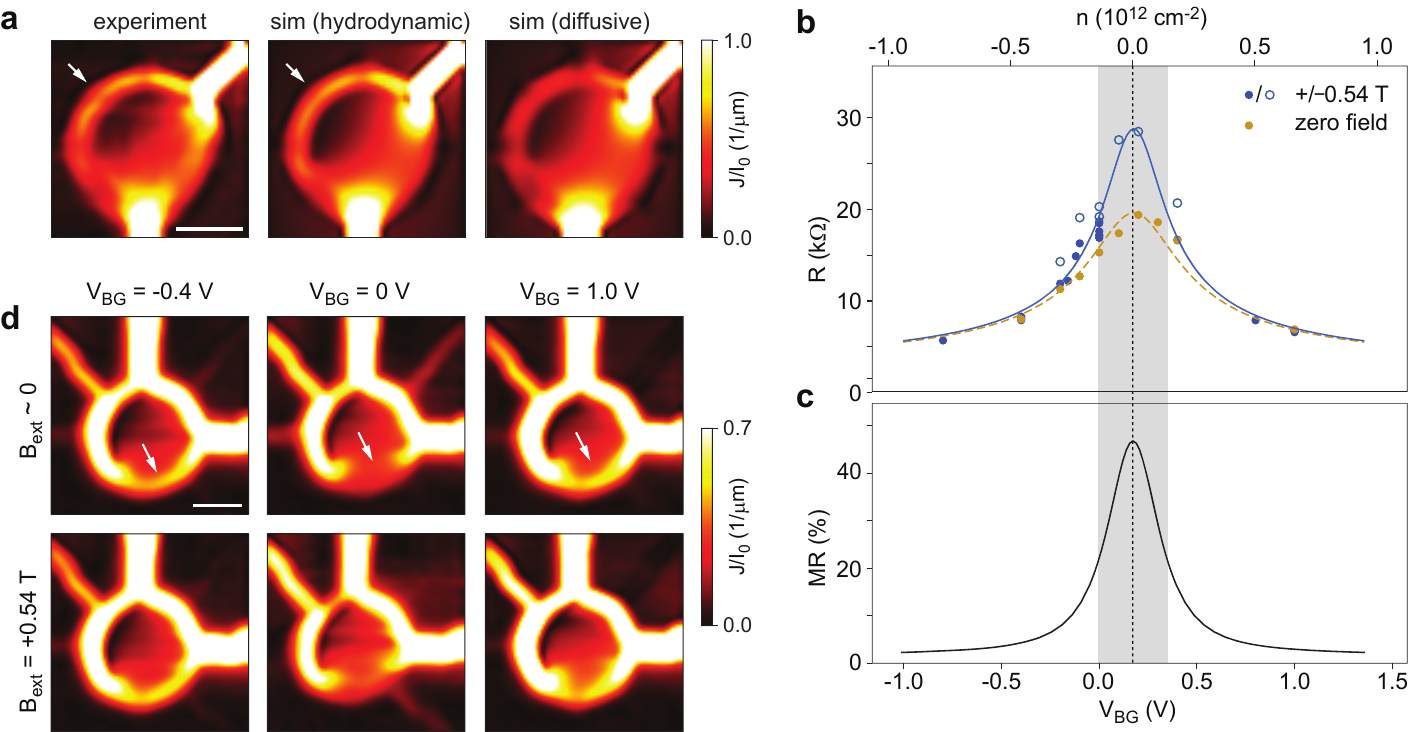}
\caption{
	\textbf{Electron hydrodynamics and carrier dependence of magnetotransport.}
	{\bf a}, Experimental data (second dataset measured at $\Bext\sim 0$) together with hydrodynamic ($\Dv=0.1\unit{\um}$) and diffusive transport simulations. The arrow indicates enhanced flow in the graphene ring for the hydrodynamic case.
	{\bf b}, Two-terminal resistance $R$ for the \textit{P-R} source-drain configuration as a function of the back-gate voltage $\Vbg$ and the corresponding carrier density $n$.  Dots are the experimental data.  Curves represent finite element simulations based on a two-carrier model~\cite{supplementary} using the parameters of Tables S1 and S2.  The CNP is at $\Vbg\approx0.175~\mr{V}$ (vertical dashed line).  The gray area shows the region of electron-hole coexistence around the Dirac cone where the Fermi energy is less than the thermal energy, $|\EF|<\kT$.
	{\bf c}, Corresponding $\MR = R(0.54\unit{T})/R(0.04\unit{T})-1$ computed from the simulation (includes $2.2\unit{k\Omega}$ \textit{P-R} contact resistance).
	{\bf d}, Measured current density maps near zero field (upper row) and at $\Bext=0.54\unit{T}$ (lower row) for hole doping ($\Vbg=-0.4\unit{V}$), near charge neutrality ($\Vbg=0$) and for electron doping ($\Vbg=1.0\unit{V}$).  The current in the graphene ring is enhanced for single carrier doping (arrows). Some current leakage occurs at contact \textit{S}. Corresponding simulated current density maps in the diffusive regime are given in Fig.~S14.  The full dataset including other $\Bext$ and $\Vbg$ is given in Figs.~S15-S17. Scale bars, $1\unit{\um}$.
	}
\label{fig4}
\end{figure*}

To investigate the role of electron hydrodynamics, we simulate the expected current distributions for diffusive ($\Dv=0$) and hydrodynamic ($\Dv=0.1\unit{\um}$, Ref.~\cite{palm2024}) transport, where $\Dv$ is the Gurzhi length~\cite{torre2015}.  We observe better qualitative agreement of the current density maps for the hydrodynamic case compared to the diffusive case (Fig.~\ref{fig4}a).  Moreover, in Fig.~\ref{fig3}e, the hydrodynamic simulation is able to explain the zero-field anomaly.  The anomaly disappears for bias fields greater than $\sim 0.1\unit{T}$ because the hydrodynamics becomes rapidly suppressed under magnetic field~\cite{berdyugin2019}.

Next, to understand why the graphene-metal hybrid displays a significant MR response (Fig.~\ref{fig1}e) even if the fraction of current in the graphene rings stays essentially constant with applied field $|\Bext|>0.1\unit{T}$ (Fig.~\ref{fig3}e), we extend measurements to the respective electron and hole-doped regimes where transport is dominated by a single carrier type.
Fig.~\ref{fig4}b shows the two-terminal resistance for various back-gate voltages between $\Vbg\approx\pm 1\unit{V}$, corresponding to a doping between approximately $n = \pm 0.8\ee{12}\unit{cm^{-2}}$ where a negative sign indicates hole carriers.  
When tuning the carrier doping of the device away from the CNP, we observe a large drop in the resistance (Fig.~\ref{fig4}b) due to the higher conductivity of doped graphene.  At the same time, $\MR\rightarrow 0\%$ at higher doping (Fig.~\ref{fig4}c).
The corresponding current density maps (Fig.~\ref{fig4}d) reveal a marked current expulsion when tuning the device away from the CNP, but little change in the current distribution for the same $n$ when applying a bias field.

Together, the data of Fig.~\ref{fig4}b-d suggest that the $\MR$ vs. $\Bext$ response near the CNP is dominated by the intrinsic MR of electron-hole coexistence in high-mobility graphene~\cite{xin2023}.  By contrast, in the single-carrier regime, the intrinsic MR of graphene is suppressed.  The geometrical contribution to $\MR$ is small for the bias fields $|\Bext|\lesssim 0.5\unit{T}$ of our study, reflected in current density maps that show only a very weak dependence on $\Bext$.  As a result, $\MR$ decreases to near-zero percent with doping.

\subsection*{Outlook}

Our work presents the first nanoscale current imaging of magnetotransport with a scanning NV magnetometer and demonstrates its value by diagnosing the mechanisms of the MR effect in a graphene-metal hybrid. In addition, our work introduces an alternative probe of the classical Hall effect based on spatial imaging of the current density. This approach is complementary to measuring the electric potential that has served as the standard method since the beginning of magnetotransport studies~\cite{hall1879}. 

Future imaging experiments will explore magnetotransport phenomena beyond the classical Hall effect. Prominent examples include the viscous Hall effect~\cite{berdyugin2019}, the anomalous Hall effect~\cite{nagaosa2010}, and the quantum Hall effect~\cite{von2020}.  Another promising playground for spatial imaging is ballistic magnetotransport~\cite{chopra2002}.  For example, ballistic MR effects related to edge and interface disorder~\cite{caridad2019}, electron-optic devices~\cite{williams2011,bhandari2016, taychatanapat2013,chen2016science}, edge scattering at graphene boundaries~\cite{tao2011}, and the reflection at superconducting interfaces~\cite{bhandari2020} are all expected to show unusual current trajectories.  Since high-mobility graphene can exhibit a mean-free path beyond $1\unit{\um}$ at room temperature~\cite{mayorov2011}, some of these phenomena could be studied under ambient condition, given the $<50\unit{nm}$ spatial resolution of the scanning NV technique~\cite{chang2017,xu2025}.  

Looking forward, nanoscale imaging of transport is especially suitable for the large class of two-dimensional materials~\cite{butler2013} prepared by exfoliation or as thin films, as they provide unobstructed access to the magnetic probe and reconstruction of spatial current patterns from magnetic stray field images is straightforward~\cite{roth1989, nowack2013,jenkins2022,aharon2022, ferguson2023, ferguson2025, dockx2025}. Bringing this powerful technique to high magnetic fields opens exciting opportunities for exploring microscopic details of magnetotransport phenomena that are difficult to probe with bulk resistance measurements.

	
\vspace{0.5cm}\textbf{Acknowledgments}

The authors thank J. Rhensius and the staff at FIRSTlab for support in nanofabriation; the Ensslin group for access to their graphene infrastructure; A. Azzani, K. Ensslin, J. Estrada Alvarez, W. Huxter, T. Ihn, and M Masseroni for helpful discussions; S. Ernst, S. Guerrero, K. Herb, P. Scheidegger, Z. Xu and T. Zhu for useful input on the magnetometer setup.
We acknowledge support from the European Research Council through ERC CoG 817720 (IMAGINE) and the Swiss National Science Foundation through Project Grant 200020\_212051.  K.W. and T.T. acknowledge support from the JSPS KAKENHI (Grant Numbers 21H05233 and 23H02052) , the CREST (JPMJCR24A5), JST and World Premier International Research Center Initiative (WPI), MEXT, Japan.


\vspace{0.5cm}\textbf{Author contributions}

M.L.P., C.D. and C.L.D. designed the experiment.
T.T. and K.W. supplied the high-quality hBN material.
C.D. fabricated the device using a stack made by M.L.P.
M.L.P., C.D. and K.K. implemented and characterized the prototypical microwave circuitry for high-bias-field configuration.
C.D. completed the high-bias-field scanning capability of the magnetometer.
C.D. carried out the experiments and performed the data analysis.
C.D. developed the theory model and performed the simulations.
C.L.D. and C.D. wrote the manuscript.
All authors discussed the results.

\section*{Materials and Methods}

\subsection*{Device fabrication}
The stack used for making the device was assembled from mechanically exfoliated flakes (graphene and hexagonal boron nitride (hBN)) through a dry transfer technique~\cite{wang2013} and annealed with a recipe reported earlier~\cite{palm2022}. Electrical contacts were defined using e-beam lithography, followed by reactive ion etching (RIE) with $\mr{CHF_3/O_2}$ plasma to remove the areas unprotected by the lithography mask. Subsequently, Cr/Au ($10/40~\mr{nm}$) was deposited through an e-beam evaporator followed by lift-off to create one-dimensional contacts to the graphene sheet~\cite{wang2013}. A second e-beam lithography step followed by RIE was used for patterning the designed device geometry. Next, the metal discs were made from Cr/Au ($7/14~\mr{nm}$) using the same recipe as for making the metal contacts. An optical microscope image and schematic cross-section of the device are shown in Fig.~S1.  A graphite back-gate was included to tune the graphene carrier density. The carrier density was estimated using a capacitance model,
\begin{align}\label{eqn: n vs Vbg}
	\Delta n(\Delta V_\mr{BG})
	   &=\varepsilon_0\varepsilon^\mr{hBN}_r\Delta V_\mr{BG}/(ed^\mr{hBN}_\mr{bot})  \nonumber \\
	   &\approx 0.8\times10^{12}~\mr{V}^{-1}\mr{cm}^{-2}\cdot\Delta V_\mr{BG}
\end{align}
where $\varepsilon^\mr{hBN}_r\approx3.76$~\cite{laturia2018} and $d^{hBN}_\mr{bot}\approx26~\mr{nm}$.

\subsection*{Scanning NV magnetometer setup}
The basic components of the setup were similar to those described in~Refs.~\cite{palm2022, palm2024}. Briefly, we used all-diamond scanning probes (QZabre Ltd.) containing a single NV center within $\sim 10\unit{nm}$ from its apex~\cite{xu2025}. Tips were made from (111) diamond material and those selected where NV centers had an out-of-plane anisotropy axis.  The tip was attached to a quartz tuning fork and operated in shear mode and with constant amplitude feedback~\cite{xu2025}.  For the optical initialization and readout of the NV centers we used a custom-built $520~\mr{nm}$ pulsed diode laser~\cite{welter2022thesis} and a single-photon counting module (Excelitas SPCM-AQRH). Manipulation of the NV spin was performed by sending microwave pulses through an aluminum bond wire located in close proximity ($\sim30~\mr{\mu m}$) to the sample.

To produce the comparably large out-of-plane bias fields needed for studying magnetotransport, we assembled a stack of cylindrical NbFeB magnets that was placed underneath the sample and positioned using a translation stage.  The maximum field at a clearance of $\sim3~\mr{mm}$ between the sample and the top of the magnet stack was approximately $\sim0.6~\mr{T}$.

Microwave pulses were generated using a microwave synthesizer (QuickSyn FSW-0020) with an output frequency range of $0.5-20.0$ GHz, an arbitrary waveform generator (Spectrum DN2.663-04), an IQ mixer (MMIQ-0218L, LO range $2.0-18.0$ GHz), and two microwave amplifiers (ZHL-16W-43-S+ and ZVE-3W-183+) used for the low frequency (LF, $1.8-4.0$ GHz) and high frequency (HF, $5.9-18.0$ GHz) bands, respectively. We used control switches to toggle between LF and HF paths.  A schematic of the circuitry is shown in Fig.~S2 and ODMR measurements up to $18\unit{GHz}$ (ca. $0.74\unit{T}$ bias field) are shown in Fig.~S3.

\subsection*{Quantum phase measurement protocol}

We used an AC quantum sensing protocol to detect the stray magnetic field generated by the currents in the device, following Ref.~\cite{palm2022}.
A schematic of the pulse protocol is shown in Fig.~S4.  Specifically, we detect a spin echo with the microwave pulses synchronized with a sinusoidal modulation of the source-drain current.  
This leads to a quantum phase $\phi$ acquired by the NV spin given by
\begin{equation}
	\phi = \gamma \frac{\pi}{2}\tau B_\mr{NV} \ ,
\end{equation}
where $\gamma=28\unit{GHz/T}$ is the gyromagnetic ratio of the NV spin, $\tau$ is the total duration of the spin-echo sequence, and $B_\mr{NV}$ is the magnetic field amplitude of the sinusoidal modulation.  The subscript indicates that the sequence is sensitive to the field component along the NV quantization axis, which in our experiments is approximately equal to the out-of-plane direction.
We extract the quantum phase using a ``four-phase'' measurement technique~\cite{palm2022} where the spin echo signal is projected along four different axes on the Bloch sphere, 
\begin{equation}
	\phi = \mr{arctan}\left(\frac{C_{3\pi/2} - C_{\pi/2}}{C_{0} - C_{\pi}}\right) \ ,
\end{equation}
where $\Phi\in\{0, \pi/2, \pi, 3\pi/2\}$ is the phase of the second $\pi/2$ microwave pulse relative to the first $\pi/2$ in the spin echo sequence, and $C_\Phi$ the measured photon count for the respective $\Phi$.

\subsection*{Differential measurement scheme}

When measuring the device at non-zero back-gate voltages $\Vbg$, we noticed the presence of back-gate leakage currents in the recorded current traces.  These manifested most prominently at $\Vsd=0$ and interfered with the source-drain signal current at $\Vsd\neq0$.  To separate the contributions from the leakage current and the source-drain current, we developed a differential detection protocol that includes source-drain voltage waveforms with opposite $\Vsd$ values at the same back-gate voltage as well as measurements with the back-gate or source-drain voltage turned off, see Fig.~S5.

Fig.~S5{\bf a} shows the recorded current traces from four sets of measurements with back-gate voltages $\{\Vbg,\Vbg,0,\Vbg\}$ and source-drain voltages $\{\Vsd,-\Vsd,0,0\}$.  For the first two measurements, the total currents are then of the form $I^{\pm}_\mr{SD} = \pm \Isig + \Ileak$, where $\Isig$ is due to the source-drain voltage and $\Ileak$ due to the back-gate voltage.  Subtraction and addition of the measurements yields $\Isig$ and $\Ileak$, respectively,
\begin{equation}\label{eqn: two-carrier model: microscopic}
	\begin{cases}
		I_\mr{sig}  &= (I^+_\mr{SD} - I^-_\mr{SD})/2 \ , \\
		I_\mr{leak} &= (I^+_\mr{SD} + I^-_\mr{SD})/2 \ .
	\end{cases}
\end{equation}
Fig.~S6 presents a collection of back-gate leakage current images as a function of $\Vbg$ for different bias magnetic fields.

\subsection*{Resistance measurements}
We measured two-terminal resistances of the device by applying an alternating source-drain voltage (amplitude $\Vsd$) with the arbitrary waveform generator (Spectrum DN2.663-04) and monitoring the device current traces. A transimpedance amplifier (FEMTO DHPCA-100) was used to amplify the current and the resulting signal was recorded using the data acquisition module of a lock-in amplifier (Zurich instruments MFLI). To characterize the carrier density dependence, we swept the back-gate voltage $\Vbg$ and used the above differential detection scheme to extract $\Io$ from $I_\mr{sig}$. The two-terminal resistance was then obtained using $R=\Vsd/\Io$. For every magnetometry scan, one current trace was recorded at each pixel and the two-terminal resistance of the scan was calculated based on the averaged current trace of all pixels.

\subsection*{Current density reconstruction}
The current density was reconstructed from the measured stray field maps using back-propagation~\cite{roth1989,chang2017}. First, the measured magnetic field projection $B_\mr{NV}(x,y)$ was transformed into Fourier space with mirrored boundary conditions to calculate the full stray field vector,
\begin{equation}
	{\hat{B}_u} = \frac{k_u}{\bm{e}\cdot\bm{k}}\BNVft  ~,u\in\{x,y,z\} \ ,
\end{equation}
where hat symbols denote 2D Fourier transformation in the image plane, and $\bm{k}=(k_x,k_y,k_z)$ is a reciprocal space vector with $k_z=-i\sqrt{k^2_x+k^2_y}$. The unit vector $\bm{e}=(e_x,e_y,e_z)=(\sin\theta\cos\varphi, \sin\theta\sin\varphi, \cos\theta)$ denotes the known anisotropy axis of the NV center, where $\theta$ and $\phi$ are the polar angle and azimuth, respectively. In our case, $\Bzft\approx\BNVft$ due to out-of-plane NV orientation in the (111) diamond probe. 

Second, the current density was computed based on the Biot-Savart law,
\begin{align}
	\Jxft &= -\frac{2}{\mu_0}e^{kz} h(k,\lambda) \hat{B}_y \ , \\
	\Jyft &= \frac{2}{\mu_0}e^{kz} h(k,\lambda) \hat{B}_x \ ,
\end{align}
where $h(k,\lambda)=\cos^2(k\lambda/4) \cdot u(2\pi/\lambda-k)$ is a Hann window function for suppressing high-frequency noise, $u(k)$ is the unit (Heaviside) step function, $z$ is the standoff distance and $\lambda$ is the cutoff wavelength of the filter. In this work, $\lambda=2z$ was used.

\subsection*{Macroscopic model of semi-classical magnetotransport in graphene}
We used a general macroscopic equation for modeling the semi-classical magnetotransport in graphene under an out-of-plane magnetic field $\bm{B}_\mr{ext}=\Bext\bm{\hat{z}}$~\cite{berdyugin2019},
\begin{equation}\label{eqn:hydro magnetotransport}
	-\sigma\bm{\nabla}\Phi(\bm{r})=(1-D^2_\nu\bm{\nabla}^2)\bm{J}(\bm{r})
	+\mu B_\mr{ext}(1+D^2_H\bm{\nabla}^2)\bm{J}(\bm{r})\times \bm{\hat{z}} \ ,
\end{equation}
where $\bm{J}(\bm{r})$ denotes the two-dimensional current density satisfying the continuity equation, 
\begin{equation}\label{eqn:continuity equation}
	\bm{\nabla}\cdot\bm{J}(\bm{r})=0 \ ,
\end{equation}
and $\Phi(\bm{r})$ denotes the electrical potential. $\sigma$ and $\mu$ correspond to the Drude conductivity and mobility (in the diffusive regime). $D_\nu$ and $D_H$ denote the diffusion length associated with the viscosity and Hall viscosity, respectively.  

We modeled the transport in graphene with a general two-carrier description,
\begin{equation}
	\bm{J}=\bm{J}_e+\bm{J}_h \ ,
\end{equation}
where $\bm{J}$ is the total current density and $\bm{J}_{e/h} = \mp en_{e/h}\bm{v}_{e/h}$ are the electron and hole current densities, respectively. $\bm{v}_{e/h}$ denote the drift velocities and $n_{e/h}$ denote the carrier densities.  The former can be obtained from the equation of motion and the latter from band theory, see Ref.~\cite{supplementary}.
Using this formalism, the conductivity and mobility of graphene are expressed by
\begin{align}\label{eqn: graphene conductivity and mobility}
	\mu_\mr{Gr} &= \frac{n_e-n_h}{n_e+n_h}\frac{\mu^2_B}{\mu_0} \ , \\
	\sigma_\mr{Gr} &= e\mu_0(n_e+n_h)\frac{1 + (\mu_\mr{Gr}\Bext)^2}{1+(\mu_B\Bext)^2} \ ,
\end{align}
where $\mu_{0}$ and $\mu_{B}$ are the zero-field and magnetotransport mobilities, respectively.  We assume electron-hole symmetry~\cite{xin2023}.

When electron density and hole density are balanced (at charge neutrality), $\mu_\mr{Gr}=0$, $\sigma_{\mr{Gr}}=e\mu_0(n_e+n_h)/(1+(\mu_B\Bext)^2)$ and therefore, the magnetoresistance becomes $\MR\propto(\mu_B\Bext)^2$. When one carrier type dominates over the other (\ie, away from the CNP), $\mu_B=\mu_0$~\cite{xin2023}. Accordingly, $\mu_\mr{Gr}=\pm\mu_0$, $\sigma_\mr{Gr}=en_{e/h}\mu_0$ and $\MR$ is zero, which recovers the formalism of the single-carrier regime.

\subsection*{Finite element simulation of current density maps}
Finite element simulations of the two-dimensional current distribution $\bm{J}(x,y)$ were performed using the \textit{Coefficient PDE} module of COMSOL Multiphysics, following previous work~\cite{aharon2022,palm2024}.
We imposed Dirichlet boundary conditions that fixed the electrical potential at source and drain contacts, specifically $\Phi_S=\Vsd$ and $\Phi_D=0$. In addition, a no-slip boundary condition fixing the current $\Jx=\Jy=0$ was imposed for the remaining boundaries of graphene.
To model the contact resistance, we included a finite graphene-metal interface region with a fixed width and an adjustable conductivity~\cite{hewett2012}.
A geometric model of the device is given in Fig.~S10. Geometry parameters and material parameters are collected in Tables~S1 and S2, respectively. 
Further details regarding the simulation are given in Ref.~\cite{supplementary}.

\subsection*{Carrier mobility estimation from resistances}
Based on the measured back-gate dependence of the two-terminal resistance together with the finite element simulation using the two-carrier model~\cite{supplementary}, the zero-field mobility of the graphene layer, $\mu_0$, can be estimated. Fig.~S11 presents the comparison between the experimental data (same as Fig.~\ref{fig4}b) and the simulated data with different $\mu_0$. As can be seen, a smaller $\mu_0$ results in a larger simulated resistance. Due to the two-terminal (source-drain) contact resistance in the actual device, the simulated resistance should not be higher than the measured resistance (at the same bias field and carrier density), which gives a lower bound for $\mu_0$ (Fig.~S11a). In addition, a larger $\mu_0$ leads to a smaller difference between the resistance at the CNP and that at high doping, which sets an upper bound of $\mu_0$ (Fig.~S11f). Overall, this estimation results in a best estimate for the carrier mobility of $\mu_0=1.35\pm0.25~\mr{m^2/V/s}$.

\subsection*{Analysis of current reorganization through current density linecuts}
To quantitatively investigate the current reorganization in the device under different experimental conditions, we analyzed the ratio between the current flowing through the metal disc and through the graphene ring at a linecut across the device. The current density across the linecut can be expressed as $J_\mr{\delta}=-J_x\sin\delta+J_y\cos\delta$, where $\delta$ denotes the tilt angle of the linecut. 

For the \textit{Q-V} configuration, we chose a linecut through the center of the device with $\delta=-32.5^{\degree}$ (see Fig.~\ref{fig3}b), which corresponds to the symmetry axis of this contact configuration. The point where $J_\mr{\delta}$ drops to zero on the right side is set to be the right boundary of the device, which then allows to define the region of the metal disc and the graphene ring  according to the device geometry (Fig.~S10).  The current passing through the graphene ring and the metal disc can then be calculated by integrating $J_\mr{\delta}$ along the linecut for the respective region. 

\subsection*{Measurement parameters for Figs.~2-4}
Fig.~2 presents the data at $\Vbg=0~\mr{V}$ ($n\approx-0.14\cdot10^{12}~\mr{cm^{-2}}$) measured with $\Io=\{17.0, 5.8, 3.7\}\unit{\uA}$ for $\{180^\circ, 135^\circ,90^\circ\}$ contact configurations.

Fig.~\ref{fig3}{\bf a} presents the data at $\Vbg=0~\mr{V}$ ($n\approx-0.14\cdot10^{12}~\mr{cm^{-2}}$) measured with $\Io=\{11.5, 5.8, 11.4\}\unit{\uA}$ for $\Bext=\{-0.53, 0.01, 0.53\}\unit{T}$. 

Fig.~\ref{fig3}{\bf c} presents the data measured with $\Io=\{11.5, 13.4, 10.4, 10.9, 11.8, 11.4\}\unit{\uA}$ for $\Bext=\{-0.53, -0.34, -0.14, 0.14, 0.35, 0.53\}\unit{T}$.
%

Fig.~\ref{fig4}{\bf d} presents data taken at $\Vbg=\{-0.4, 0, 1.0\}\unit{V}$ ($n\approx\{-0.46, -0.14, 0.66\}\cdot10^{12}~\mr{cm^{-2}}$). $\Io=5-27\unit{\uA}$.

For the current reconstruction, $z=100~\mr{nm}$ was used for data measured with \{\textit{R-T}, \textit{R-V}, \textit{Q-V}\} configurations (Fig.~\ref{fig2} and Fig.~\ref{fig3}), and $z=125~\mr{nm}$ was used for data measured with the \textit{P-R} configuration (Fig.~\ref{fig4}). The standoff distance was determined from the magnetic field profile of the current-carrying graphene and represents the vertical distance between the NV probe and the graphene sheet.  The cutoff wavelength of the Hann filter was set to be $\lambda=2z$. $\theta=3\degree$ and $\varphi=35\degree$ were used as NV angles. All measurements were acquired with the same scanning probe.

\input{"references.bbl"}

\end{document}

%% file: references.bbl
%

%% file: manuscript.bbl
\begin{thebibliography}{47}%
\makeatletter
\providecommand \@ifxundefined [1]{%
 \@ifx{#1\undefined}
}%
\providecommand \@ifnum [1]{%
 \ifnum #1\expandafter \@firstoftwo
 \else \expandafter \@secondoftwo
 \fi
}%
\providecommand \@ifx [1]{%
 \ifx #1\expandafter \@firstoftwo
 \else \expandafter \@secondoftwo
 \fi
}%
\providecommand \natexlab [1]{#1}%
\providecommand \enquote  [1]{``#1''}%
\providecommand \bibnamefont  [1]{#1}%
\providecommand \bibfnamefont [1]{#1}%
\providecommand \citenamefont [1]{#1}%
\providecommand \href@noop [0]{\@secondoftwo}%
\providecommand \href [0]{\begingroup \@sanitize@url \@href}%
\providecommand \@href[1]{\@@startlink{#1}\@@href}%
\providecommand \@@href[1]{\endgroup#1\@@endlink}%
\providecommand \@sanitize@url [0]{\catcode `\\12\catcode `\$12\catcode
  `\&12\catcode `\#12\catcode `\^12\catcode `\_12\catcode `\%12\relax}%
\providecommand \@@startlink[1]{}%
\providecommand \@@endlink[0]{}%
\providecommand \url  [0]{\begingroup\@sanitize@url \@url }%
\providecommand \@url [1]{\endgroup\@href {#1}{\urlprefix }}%
\providecommand \urlprefix  [0]{URL }%
\providecommand \Eprint [0]{\href }%
\providecommand \doibase [0]{https://doi.org/}%
\providecommand \selectlanguage [0]{\@gobble}%
\providecommand \bibinfo  [0]{\@secondoftwo}%
\providecommand \bibfield  [0]{\@secondoftwo}%
\providecommand \translation [1]{[#1]}%
\providecommand \BibitemOpen [0]{}%
\providecommand \bibitemStop [0]{}%
\providecommand \bibitemNoStop [0]{.\EOS\space}%
\providecommand \EOS [0]{\spacefactor3000\relax}%
\providecommand \BibitemShut  [1]{\csname bibitem#1\endcsname}%
\let\auto@bib@innerbib\@empty
\bibitem [{\citenamefont {Hall}\ \emph {et~al.}(1879)\citenamefont {Hall} \emph
  {et~al.}}]{hall1879}%
  \BibitemOpen
  \bibfield  {author} {\bibinfo {author} {\bibfnamefont {E.~H.}\ \bibnamefont
  {Hall}} \emph {et~al.},\ }\bibfield  {title} {\bibinfo {title} {On a new
  action of the magnet on electric currents},\ }\href@noop {} {\bibfield
  {journal} {\bibinfo  {journal} {American Journal of Mathematics}\ }\textbf
  {\bibinfo {volume} {2}},\ \bibinfo {pages} {287} (\bibinfo {year}
  {1879})}\BibitemShut {NoStop}%
\bibitem [{\citenamefont {Solin}\ \emph {et~al.}(2000)\citenamefont {Solin},
  \citenamefont {Thio}, \citenamefont {Hines},\ and\ \citenamefont
  {Heremans}}]{solin2000}%
  \BibitemOpen
  \bibfield  {author} {\bibinfo {author} {\bibfnamefont {S.}~\bibnamefont
  {Solin}}, \bibinfo {author} {\bibfnamefont {T.}~\bibnamefont {Thio}},
  \bibinfo {author} {\bibfnamefont {D.}~\bibnamefont {Hines}},\ and\ \bibinfo
  {author} {\bibfnamefont {J.}~\bibnamefont {Heremans}},\ }\bibfield  {title}
  {\bibinfo {title} {Enhanced room-temperature geometric magnetoresistance in
  inhomogeneous narrow-gap semiconductors},\ }\href@noop {} {\bibfield
  {journal} {\bibinfo  {journal} {Science}\ }\textbf {\bibinfo {volume}
  {289}},\ \bibinfo {pages} {1530} (\bibinfo {year} {2000})}\BibitemShut
  {NoStop}%
\bibitem [{\citenamefont {Pomar}\ \emph
  {et~al.}(2024{\natexlab{a}})\citenamefont {Pomar}, \citenamefont {Erlandsen},
  \citenamefont {Zhou}, \citenamefont {Iliushyn}, \citenamefont {Bjork},\ and\
  \citenamefont {Christensen}}]{pomar2024amt}%
  \BibitemOpen
  \bibfield  {author} {\bibinfo {author} {\bibfnamefont {T.~D.}\ \bibnamefont
  {Pomar}}, \bibinfo {author} {\bibfnamefont {R.}~\bibnamefont {Erlandsen}},
  \bibinfo {author} {\bibfnamefont {B.}~\bibnamefont {Zhou}}, \bibinfo {author}
  {\bibfnamefont {L.}~\bibnamefont {Iliushyn}}, \bibinfo {author}
  {\bibfnamefont {R.}~\bibnamefont {Bjork}},\ and\ \bibinfo {author}
  {\bibfnamefont {D.~V.}\ \bibnamefont {Christensen}},\ }\bibfield  {title}
  {\bibinfo {title} {Extraordinary magnetometry: a review on extraordinary
  magnetoresistance},\ }\href {https://doi.org/10.1016/j.apmt.2024.102219}
  {\bibfield  {journal} {\bibinfo  {journal} {Applied Materials Today}\
  }\textbf {\bibinfo {volume} {38}},\ \bibinfo {pages} {102219} (\bibinfo
  {year} {2024}{\natexlab{a}})}\BibitemShut {NoStop}%
\bibitem [{\citenamefont {Ku}\ \emph {et~al.}(2020)\citenamefont {Ku},
  \citenamefont {Zhou}, \citenamefont {Li}, \citenamefont {Shin}, \citenamefont
  {Shi}, \citenamefont {Burch}, \citenamefont {Anderson}, \citenamefont
  {Pierce}, \citenamefont {Xie}, \citenamefont {Hamo}, \citenamefont {Vool},
  \citenamefont {Zhang}, \citenamefont {Casola}, \citenamefont {Taniguchi},
  \citenamefont {Watanabe}, \citenamefont {Fogler}, \citenamefont {Kim},
  \citenamefont {Yacoby},\ and\ \citenamefont {Walsworth}}]{ku2020}%
  \BibitemOpen
  \bibfield  {author} {\bibinfo {author} {\bibfnamefont {M.~J.~H.}\
  \bibnamefont {Ku}}, \bibinfo {author} {\bibfnamefont {T.~X.}\ \bibnamefont
  {Zhou}}, \bibinfo {author} {\bibfnamefont {Q.}~\bibnamefont {Li}}, \bibinfo
  {author} {\bibfnamefont {Y.~J.}\ \bibnamefont {Shin}}, \bibinfo {author}
  {\bibfnamefont {J.~K.}\ \bibnamefont {Shi}}, \bibinfo {author} {\bibfnamefont
  {C.}~\bibnamefont {Burch}}, \bibinfo {author} {\bibfnamefont {L.~E.}\
  \bibnamefont {Anderson}}, \bibinfo {author} {\bibfnamefont {A.~T.}\
  \bibnamefont {Pierce}}, \bibinfo {author} {\bibfnamefont {Y.}~\bibnamefont
  {Xie}}, \bibinfo {author} {\bibfnamefont {A.}~\bibnamefont {Hamo}}, \bibinfo
  {author} {\bibfnamefont {U.}~\bibnamefont {Vool}}, \bibinfo {author}
  {\bibfnamefont {H.}~\bibnamefont {Zhang}}, \bibinfo {author} {\bibfnamefont
  {F.}~\bibnamefont {Casola}}, \bibinfo {author} {\bibfnamefont
  {T.}~\bibnamefont {Taniguchi}}, \bibinfo {author} {\bibfnamefont
  {K.}~\bibnamefont {Watanabe}}, \bibinfo {author} {\bibfnamefont {M.~M.}\
  \bibnamefont {Fogler}}, \bibinfo {author} {\bibfnamefont {P.}~\bibnamefont
  {Kim}}, \bibinfo {author} {\bibfnamefont {A.}~\bibnamefont {Yacoby}},\ and\
  \bibinfo {author} {\bibfnamefont {R.~L.}\ \bibnamefont {Walsworth}},\
  }\bibfield  {title} {\bibinfo {title} {Imaging viscous flow of the {{Dirac}}
  fluid in graphene},\ }\href {https://doi.org/10.1038/s41586-020-2507-2}
  {\bibfield  {journal} {\bibinfo  {journal} {Nature}\ }\textbf {\bibinfo
  {volume} {583}},\ \bibinfo {pages} {537} (\bibinfo {year}
  {2020})}\BibitemShut {NoStop}%
\bibitem [{\citenamefont {Vool}\ \emph {et~al.}(2021)\citenamefont {Vool},
  \citenamefont {Hamo}, \citenamefont {Varnavides}, \citenamefont {Wang},
  \citenamefont {Zhou}, \citenamefont {Kumar}, \citenamefont {Dovzhenko},
  \citenamefont {Qiu}, \citenamefont {Garcia}, \citenamefont {Pierce},
  \citenamefont {Gooth}, \citenamefont {Anikeeva}, \citenamefont {Felser},
  \citenamefont {Narang},\ and\ \citenamefont {Yacoby}}]{vool2021}%
  \BibitemOpen
  \bibfield  {author} {\bibinfo {author} {\bibfnamefont {U.}~\bibnamefont
  {Vool}}, \bibinfo {author} {\bibfnamefont {A.}~\bibnamefont {Hamo}}, \bibinfo
  {author} {\bibfnamefont {G.}~\bibnamefont {Varnavides}}, \bibinfo {author}
  {\bibfnamefont {Y.}~\bibnamefont {Wang}}, \bibinfo {author} {\bibfnamefont
  {T.~X.}\ \bibnamefont {Zhou}}, \bibinfo {author} {\bibfnamefont
  {N.}~\bibnamefont {Kumar}}, \bibinfo {author} {\bibfnamefont
  {Y.}~\bibnamefont {Dovzhenko}}, \bibinfo {author} {\bibfnamefont
  {Z.}~\bibnamefont {Qiu}}, \bibinfo {author} {\bibfnamefont {C.~A.~C.}\
  \bibnamefont {Garcia}}, \bibinfo {author} {\bibfnamefont {A.~T.}\
  \bibnamefont {Pierce}}, \bibinfo {author} {\bibfnamefont {J.}~\bibnamefont
  {Gooth}}, \bibinfo {author} {\bibfnamefont {P.}~\bibnamefont {Anikeeva}},
  \bibinfo {author} {\bibfnamefont {C.}~\bibnamefont {Felser}}, \bibinfo
  {author} {\bibfnamefont {P.}~\bibnamefont {Narang}},\ and\ \bibinfo {author}
  {\bibfnamefont {A.}~\bibnamefont {Yacoby}},\ }\bibfield  {title} {\bibinfo
  {title} {Imaging phonon-mediated hydrodynamic flow in {WTe}$_2$},\ }\href
  {https://doi.org/10.1038/s41567-021-01341-w} {\bibfield  {journal} {\bibinfo
  {journal} {Nature Physics}\ }\textbf {\bibinfo {volume} {17}},\ \bibinfo
  {pages} {1216} (\bibinfo {year} {2021})}\BibitemShut {NoStop}%
\bibitem [{\citenamefont {Palm}\ \emph {et~al.}(2022)\citenamefont {Palm},
  \citenamefont {Huxter}, \citenamefont {Welter}, \citenamefont {Ernst},
  \citenamefont {Scheidegger}, \citenamefont {Diesch}, \citenamefont {Chang},
  \citenamefont {Rickhaus}, \citenamefont {Taniguchi}, \citenamefont
  {Watanabe}, \citenamefont {Ensslin},\ and\ \citenamefont {Degen}}]{palm2022}%
  \BibitemOpen
  \bibfield  {author} {\bibinfo {author} {\bibfnamefont {M.~L.}\ \bibnamefont
  {Palm}}, \bibinfo {author} {\bibfnamefont {W.~S.}\ \bibnamefont {Huxter}},
  \bibinfo {author} {\bibfnamefont {P.}~\bibnamefont {Welter}}, \bibinfo
  {author} {\bibfnamefont {S.}~\bibnamefont {Ernst}}, \bibinfo {author}
  {\bibfnamefont {P.~J.}\ \bibnamefont {Scheidegger}}, \bibinfo {author}
  {\bibfnamefont {S.}~\bibnamefont {Diesch}}, \bibinfo {author} {\bibfnamefont
  {K.}~\bibnamefont {Chang}}, \bibinfo {author} {\bibfnamefont
  {P.}~\bibnamefont {Rickhaus}}, \bibinfo {author} {\bibfnamefont
  {T.}~\bibnamefont {Taniguchi}}, \bibinfo {author} {\bibfnamefont
  {K.}~\bibnamefont {Watanabe}}, \bibinfo {author} {\bibfnamefont
  {K.}~\bibnamefont {Ensslin}},\ and\ \bibinfo {author} {\bibfnamefont {C.~L.}\
  \bibnamefont {Degen}},\ }\bibfield  {title} {\bibinfo {title} {Imaging of
  submicroampere currents in bilayer graphene using a scanning diamond
  magnetometer},\ }\bibfield  {journal} {\bibinfo  {journal} {Physical Review
  Applied}\ }\textbf {\bibinfo {volume} {17}},\ \href
  {https://doi.org/10.1103/physrevapplied.17.054008}
  {10.1103/physrevapplied.17.054008} (\bibinfo {year} {2022})\BibitemShut
  {NoStop}%
\bibitem [{\citenamefont {Palm}\ \emph {et~al.}(2024)\citenamefont {Palm},
  \citenamefont {Ding}, \citenamefont {Huxter}, \citenamefont {Taniguchi},
  \citenamefont {Watanabe},\ and\ \citenamefont {Degen}}]{palm2024}%
  \BibitemOpen
  \bibfield  {author} {\bibinfo {author} {\bibfnamefont {M.~L.}\ \bibnamefont
  {Palm}}, \bibinfo {author} {\bibfnamefont {C.}~\bibnamefont {Ding}}, \bibinfo
  {author} {\bibfnamefont {W.~S.}\ \bibnamefont {Huxter}}, \bibinfo {author}
  {\bibfnamefont {T.}~\bibnamefont {Taniguchi}}, \bibinfo {author}
  {\bibfnamefont {K.}~\bibnamefont {Watanabe}},\ and\ \bibinfo {author}
  {\bibfnamefont {C.~L.}\ \bibnamefont {Degen}},\ }\bibfield  {title} {\bibinfo
  {title} {Observation of current whirlpools in graphene at room temperature},\
  }\href {https://doi.org/10.1126/science.adj2167} {\bibfield  {journal}
  {\bibinfo  {journal} {Science}\ }\textbf {\bibinfo {volume} {384}},\ \bibinfo
  {pages} {465} (\bibinfo {year} {2024})}\BibitemShut {NoStop}%
\bibitem [{sup()}]{supplementary}%
  \BibitemOpen
  \href@noop {} {\bibinfo  {journal} {See Supplementary Materials accompanying
  this manuscript}\ }\BibitemShut {NoStop}%
\bibitem [{\citenamefont {Lu}\ \emph {et~al.}(2011)\citenamefont {Lu},
  \citenamefont {Zhang}, \citenamefont {Shi}, \citenamefont {Wang},
  \citenamefont {Zheng}, \citenamefont {Zhang}, \citenamefont {Wang},
  \citenamefont {Tang},\ and\ \citenamefont {Sheng}}]{lu2011}%
  \BibitemOpen
\bibfield  {journal} {  }\bibfield  {author} {\bibinfo {author} {\bibfnamefont
  {J.}~\bibnamefont {Lu}}, \bibinfo {author} {\bibfnamefont {H.}~\bibnamefont
  {Zhang}}, \bibinfo {author} {\bibfnamefont {W.}~\bibnamefont {Shi}}, \bibinfo
  {author} {\bibfnamefont {Z.}~\bibnamefont {Wang}}, \bibinfo {author}
  {\bibfnamefont {Y.}~\bibnamefont {Zheng}}, \bibinfo {author} {\bibfnamefont
  {T.}~\bibnamefont {Zhang}}, \bibinfo {author} {\bibfnamefont
  {N.}~\bibnamefont {Wang}}, \bibinfo {author} {\bibfnamefont {Z.}~\bibnamefont
  {Tang}},\ and\ \bibinfo {author} {\bibfnamefont {P.}~\bibnamefont {Sheng}},\
  }\bibfield  {title} {\bibinfo {title} {Graphene magnetoresistance device in
  van der pauw geometry},\ }\href@noop {} {\bibfield  {journal} {\bibinfo
  {journal} {Nano letters}\ }\textbf {\bibinfo {volume} {11}},\ \bibinfo
  {pages} {2973} (\bibinfo {year} {2011})}\BibitemShut {NoStop}%
\bibitem [{\citenamefont {Zhou}\ \emph {et~al.}(2020)\citenamefont {Zhou},
  \citenamefont {Watanabe}, \citenamefont {Taniguchi},\ and\ \citenamefont
  {Henriksen}}]{zhou2020apl}%
  \BibitemOpen
  \bibfield  {author} {\bibinfo {author} {\bibfnamefont {B.}~\bibnamefont
  {Zhou}}, \bibinfo {author} {\bibfnamefont {K.}~\bibnamefont {Watanabe}},
  \bibinfo {author} {\bibfnamefont {T.}~\bibnamefont {Taniguchi}},\ and\
  \bibinfo {author} {\bibfnamefont {E.~A.}\ \bibnamefont {Henriksen}},\
  }\bibfield  {title} {\bibinfo {title} {Extraordinary magnetoresistance in
  encapsulated monolayer graphene devices},\ }\bibfield  {journal} {\bibinfo
  {journal} {Applied Physics Letters}\ }\textbf {\bibinfo {volume} {116}},\
  \href {https://doi.org/10.1063/1.5142021} {10.1063/1.5142021} (\bibinfo
  {year} {2020})\BibitemShut {NoStop}%
\bibitem [{\citenamefont {Rohner}\ \emph {et~al.}(2019)\citenamefont {Rohner},
  \citenamefont {Happacher}, \citenamefont {Reiser}, \citenamefont {Tschudin},
  \citenamefont {Tallaire}, \citenamefont {Achard}, \citenamefont {Shields},\
  and\ \citenamefont {Maletinsky}}]{rohner2019}%
  \BibitemOpen
  \bibfield  {author} {\bibinfo {author} {\bibfnamefont {D.}~\bibnamefont
  {Rohner}}, \bibinfo {author} {\bibfnamefont {J.}~\bibnamefont {Happacher}},
  \bibinfo {author} {\bibfnamefont {P.}~\bibnamefont {Reiser}}, \bibinfo
  {author} {\bibfnamefont {M.~A.}\ \bibnamefont {Tschudin}}, \bibinfo {author}
  {\bibfnamefont {A.}~\bibnamefont {Tallaire}}, \bibinfo {author}
  {\bibfnamefont {J.}~\bibnamefont {Achard}}, \bibinfo {author} {\bibfnamefont
  {B.~J.}\ \bibnamefont {Shields}},\ and\ \bibinfo {author} {\bibfnamefont
  {P.}~\bibnamefont {Maletinsky}},\ }\bibfield  {title} {\bibinfo {title}
  {(111)-oriented single crystal diamond tips for nanoscale scanning probe
  imaging of out-of-plane magnetic fields},\ }\href
  {https://doi.org/10.1063/1.5127101} {\bibfield  {journal} {\bibinfo
  {journal} {Appl. Phys. Lett.}\ }\textbf {\bibinfo {volume} {115}},\ \bibinfo
  {pages} {192401} (\bibinfo {year} {2019})}\BibitemShut {NoStop}%
\bibitem [{\citenamefont {Epstein}\ \emph {et~al.}(2005)\citenamefont
  {Epstein}, \citenamefont {Mendoza}, \citenamefont {Kato},\ and\ \citenamefont
  {Awschalom}}]{epstein2005}%
  \BibitemOpen
  \bibfield  {author} {\bibinfo {author} {\bibfnamefont {R.~J.}\ \bibnamefont
  {Epstein}}, \bibinfo {author} {\bibfnamefont {F.~M.}\ \bibnamefont
  {Mendoza}}, \bibinfo {author} {\bibfnamefont {Y.~K.}\ \bibnamefont {Kato}},\
  and\ \bibinfo {author} {\bibfnamefont {D.~D.}\ \bibnamefont {Awschalom}},\
  }\bibfield  {title} {\bibinfo {title} {Anisotropic interactions of a single
  spin and dark-spin spectroscopy in diamond},\ }\href
  {https://doi.org/10.1038/nphys141} {\bibfield  {journal} {\bibinfo  {journal}
  {Nat. Phys.}\ }\textbf {\bibinfo {volume} {1}},\ \bibinfo {eid} {94}
  (\bibinfo {year} {2005})}\BibitemShut {NoStop}%
\bibitem [{\citenamefont {Tetienne}\ \emph {et~al.}(2012)\citenamefont
  {Tetienne}, \citenamefont {Rondin}, \citenamefont {Spinicelli}, \citenamefont
  {Chipaux}, \citenamefont {Debuisschert}, \citenamefont {Roch},\ and\
  \citenamefont {Jacques}}]{tetienne2012}%
  \BibitemOpen
  \bibfield  {author} {\bibinfo {author} {\bibfnamefont {J.}~\bibnamefont
  {Tetienne}}, \bibinfo {author} {\bibfnamefont {L.}~\bibnamefont {Rondin}},
  \bibinfo {author} {\bibfnamefont {P.}~\bibnamefont {Spinicelli}}, \bibinfo
  {author} {\bibfnamefont {M.}~\bibnamefont {Chipaux}}, \bibinfo {author}
  {\bibfnamefont {T.}~\bibnamefont {Debuisschert}}, \bibinfo {author}
  {\bibfnamefont {J.}~\bibnamefont {Roch}},\ and\ \bibinfo {author}
  {\bibfnamefont {V.}~\bibnamefont {Jacques}},\ }\bibfield  {title} {\bibinfo
  {title} {Magnetic-field-dependent photodynamics of single {{NV}} defects in
  diamond: an application to qualitative all-optical magnetic imaging},\ }\href
  {http://dx.doi.org/10.1088/1367-2630/14/10/103033} {\bibfield  {journal}
  {\bibinfo  {journal} {New Journal of Physics}\ }\textbf {\bibinfo {volume}
  {14}},\ \bibinfo {pages} {103033} (\bibinfo {year} {2012})}\BibitemShut
  {NoStop}%
\bibitem [{\citenamefont {Neto}\ \emph {et~al.}(2009)\citenamefont {Neto},
  \citenamefont {Guinea}, \citenamefont {Peres}, \citenamefont {Novoselov},\
  and\ \citenamefont {Geim}}]{castroneto2009}%
  \BibitemOpen
  \bibfield  {author} {\bibinfo {author} {\bibfnamefont {A.~C.}\ \bibnamefont
  {Neto}}, \bibinfo {author} {\bibfnamefont {F.}~\bibnamefont {Guinea}},
  \bibinfo {author} {\bibfnamefont {N.}~\bibnamefont {Peres}}, \bibinfo
  {author} {\bibfnamefont {K.}~\bibnamefont {Novoselov}},\ and\ \bibinfo
  {author} {\bibfnamefont {A.}~\bibnamefont {Geim}},\ }\bibfield  {title}
  {\bibinfo {title} {The electronic properties of graphene},\ }\href
  {https://doi.org/10.1103/RevModPhys.81.109} {\bibfield  {journal} {\bibinfo
  {journal} {Rev. Mod. Phys.}\ }\textbf {\bibinfo {volume} {81}},\ \bibinfo
  {pages} {109} (\bibinfo {year} {2009})}\BibitemShut {NoStop}%
\bibitem [{\citenamefont {Roth}\ \emph {et~al.}(1989)\citenamefont {Roth},
  \citenamefont {Sepulveda},\ and\ \citenamefont {Wikswo}}]{roth1989}%
  \BibitemOpen
  \bibfield  {author} {\bibinfo {author} {\bibfnamefont {B.~J.}\ \bibnamefont
  {Roth}}, \bibinfo {author} {\bibfnamefont {N.~G.}\ \bibnamefont
  {Sepulveda}},\ and\ \bibinfo {author} {\bibfnamefont {J.~P.}\ \bibnamefont
  {Wikswo}},\ }\bibfield  {title} {\bibinfo {title} {Using a magnetometer to
  image a two-dimensional current distribution},\ }\href
  {https://doi.org/10.1063/1.342549} {\bibfield  {journal} {\bibinfo  {journal}
  {J. Appl. Phys.}\ }\textbf {\bibinfo {volume} {65}},\ \bibinfo {pages} {361}
  (\bibinfo {year} {1989})}\BibitemShut {NoStop}%
\bibitem [{\citenamefont {Chang}\ \emph {et~al.}(2017)\citenamefont {Chang},
  \citenamefont {Eichler}, \citenamefont {Rhensius}, \citenamefont
  {Lorenzelli},\ and\ \citenamefont {Degen}}]{chang2017}%
  \BibitemOpen
  \bibfield  {author} {\bibinfo {author} {\bibfnamefont {K.}~\bibnamefont
  {Chang}}, \bibinfo {author} {\bibfnamefont {A.}~\bibnamefont {Eichler}},
  \bibinfo {author} {\bibfnamefont {J.}~\bibnamefont {Rhensius}}, \bibinfo
  {author} {\bibfnamefont {L.}~\bibnamefont {Lorenzelli}},\ and\ \bibinfo
  {author} {\bibfnamefont {C.~L.}\ \bibnamefont {Degen}},\ }\bibfield  {title}
  {\bibinfo {title} {Nanoscale imaging of current density with a single-spin
  magnetometer},\ }\href {https://doi.org/10.1021/acs.nanolett.6b05304}
  {\bibfield  {journal} {\bibinfo  {journal} {Nano Letters}\ }\textbf {\bibinfo
  {volume} {17}},\ \bibinfo {pages} {2367} (\bibinfo {year}
  {2017})}\BibitemShut {NoStop}%
\bibitem [{\citenamefont {Holz}\ \emph {et~al.}(2003)\citenamefont {Holz},
  \citenamefont {Kronenwerth},\ and\ \citenamefont {Grundler}}]{holz2003}%
  \BibitemOpen
  \bibfield  {author} {\bibinfo {author} {\bibfnamefont {M.}~\bibnamefont
  {Holz}}, \bibinfo {author} {\bibfnamefont {O.}~\bibnamefont {Kronenwerth}},\
  and\ \bibinfo {author} {\bibfnamefont {D.}~\bibnamefont {Grundler}},\
  }\bibfield  {title} {\bibinfo {title} {Magnetoresistance of
  semiconductor-metal hybrid structures: the effects of material parameters and
  contact resistance},\ }\href {https://doi.org/10.1103/PhysRevB.67.195312}
  {\bibfield  {journal} {\bibinfo  {journal} {Physical Review B}\ }\textbf
  {\bibinfo {volume} {67}},\ \bibinfo {pages} {195312} (\bibinfo {year}
  {2003})}\BibitemShut {NoStop}%
\bibitem [{\citenamefont {Pomar}\ \emph
  {et~al.}(2024{\natexlab{b}})\citenamefont {Pomar}, \citenamefont
  {Steegemans}, \citenamefont {Kumar}, \citenamefont {Bjork}, \citenamefont
  {Lei}, \citenamefont {Cheah}, \citenamefont {Schott}, \citenamefont {Bogild},
  \citenamefont {Pryds}, \citenamefont {Wegscheider},\ and\ \citenamefont
  {Christensen}}]{pomar2024arxiv}%
  \BibitemOpen
  \bibfield  {author} {\bibinfo {author} {\bibfnamefont {T.~D.}\ \bibnamefont
  {Pomar}}, \bibinfo {author} {\bibfnamefont {T.}~\bibnamefont {Steegemans}},
  \bibinfo {author} {\bibfnamefont {S.}~\bibnamefont {Kumar}}, \bibinfo
  {author} {\bibfnamefont {R.}~\bibnamefont {Bjork}}, \bibinfo {author}
  {\bibfnamefont {Z.}~\bibnamefont {Lei}}, \bibinfo {author} {\bibfnamefont
  {E.}~\bibnamefont {Cheah}}, \bibinfo {author} {\bibfnamefont
  {R.}~\bibnamefont {Schott}}, \bibinfo {author} {\bibfnamefont
  {P.}~\bibnamefont {Bogild}}, \bibinfo {author} {\bibfnamefont
  {N.}~\bibnamefont {Pryds}}, \bibinfo {author} {\bibfnamefont
  {W.}~\bibnamefont {Wegscheider}},\ and\ \bibinfo {author} {\bibfnamefont
  {D.~V.}\ \bibnamefont {Christensen}},\ }\bibfield  {title} {\bibinfo {title}
  {Improving electrical contact quality and extraordinary magnetoresistance in
  high mobility iii-v semiconductors},\ }\href
  {https://arxiv.org/abs/2410.17713} {\bibfield  {journal} {\bibinfo  {journal}
  {arXiv:2410.17713}\ } (\bibinfo {year} {2024}{\natexlab{b}})}\BibitemShut
  {NoStop}%
\bibitem [{\citenamefont {Wang}\ \emph {et~al.}(2013)\citenamefont {Wang},
  \citenamefont {Meric}, \citenamefont {Huang}, \citenamefont {Gao},
  \citenamefont {Gao}, \citenamefont {Tran}, \citenamefont {Taniguchi},
  \citenamefont {Watanabe}, \citenamefont {Campos}, \citenamefont {Muller},
  \citenamefont {Guo}, \citenamefont {Kim}, \citenamefont {Hone}, \citenamefont
  {Shepard},\ and\ \citenamefont {Dean}}]{wang2013}%
  \BibitemOpen
  \bibfield  {author} {\bibinfo {author} {\bibfnamefont {L.}~\bibnamefont
  {Wang}}, \bibinfo {author} {\bibfnamefont {I.}~\bibnamefont {Meric}},
  \bibinfo {author} {\bibfnamefont {P.~Y.}\ \bibnamefont {Huang}}, \bibinfo
  {author} {\bibfnamefont {Q.}~\bibnamefont {Gao}}, \bibinfo {author}
  {\bibfnamefont {Y.}~\bibnamefont {Gao}}, \bibinfo {author} {\bibfnamefont
  {H.}~\bibnamefont {Tran}}, \bibinfo {author} {\bibfnamefont {T.}~\bibnamefont
  {Taniguchi}}, \bibinfo {author} {\bibfnamefont {K.}~\bibnamefont {Watanabe}},
  \bibinfo {author} {\bibfnamefont {L.~M.}\ \bibnamefont {Campos}}, \bibinfo
  {author} {\bibfnamefont {D.~A.}\ \bibnamefont {Muller}}, \bibinfo {author}
  {\bibfnamefont {J.}~\bibnamefont {Guo}}, \bibinfo {author} {\bibfnamefont
  {P.}~\bibnamefont {Kim}}, \bibinfo {author} {\bibfnamefont {J.}~\bibnamefont
  {Hone}}, \bibinfo {author} {\bibfnamefont {K.~L.}\ \bibnamefont {Shepard}},\
  and\ \bibinfo {author} {\bibfnamefont {C.~R.}\ \bibnamefont {Dean}},\
  }\bibfield  {title} {\bibinfo {title} {One-dimensional electrical contact to
  a two-dimensional material},\ }\href
  {https://doi.org/10.1126/science.1244358} {\bibfield  {journal} {\bibinfo
  {journal} {Science}\ }\textbf {\bibinfo {volume} {342}},\ \bibinfo {pages}
  {614} (\bibinfo {year} {2013})}\BibitemShut {NoStop}%
\bibitem [{\citenamefont {Kwon}\ \emph {et~al.}(2022)\citenamefont {Kwon},
  \citenamefont {Choi}, \citenamefont {Lee}, \citenamefont {Kim}, \citenamefont
  {Jeong}, \citenamefont {Jeong}, \citenamefont {Baik}, \citenamefont {Kwon},
  \citenamefont {Ahn}, \citenamefont {Lee} \emph
  {et~al.}}]{kwon2022interaction}%
  \BibitemOpen
  \bibfield  {author} {\bibinfo {author} {\bibfnamefont {G.}~\bibnamefont
  {Kwon}}, \bibinfo {author} {\bibfnamefont {Y.-H.}\ \bibnamefont {Choi}},
  \bibinfo {author} {\bibfnamefont {H.}~\bibnamefont {Lee}}, \bibinfo {author}
  {\bibfnamefont {H.-S.}\ \bibnamefont {Kim}}, \bibinfo {author} {\bibfnamefont
  {J.}~\bibnamefont {Jeong}}, \bibinfo {author} {\bibfnamefont
  {K.}~\bibnamefont {Jeong}}, \bibinfo {author} {\bibfnamefont
  {M.}~\bibnamefont {Baik}}, \bibinfo {author} {\bibfnamefont {H.}~\bibnamefont
  {Kwon}}, \bibinfo {author} {\bibfnamefont {J.}~\bibnamefont {Ahn}}, \bibinfo
  {author} {\bibfnamefont {E.}~\bibnamefont {Lee}}, \emph {et~al.},\ }\bibfield
   {title} {\bibinfo {title} {Interaction-and defect-free van der waals
  contacts between metals and two-dimensional semiconductors},\ }\href@noop {}
  {\bibfield  {journal} {\bibinfo  {journal} {Nature Electronics}\ }\textbf
  {\bibinfo {volume} {5}},\ \bibinfo {pages} {241} (\bibinfo {year}
  {2022})}\BibitemShut {NoStop}%
\bibitem [{\citenamefont {Moller}\ \emph {et~al.}(2003)\citenamefont {Moller},
  \citenamefont {Grundler}, \citenamefont {Kronenwerth}, \citenamefont {Heyn},\
  and\ \citenamefont {Heitmann}}]{moller2003}%
  \BibitemOpen
  \bibfield  {author} {\bibinfo {author} {\bibfnamefont {C.~H.}\ \bibnamefont
  {Moller}}, \bibinfo {author} {\bibfnamefont {D.}~\bibnamefont {Grundler}},
  \bibinfo {author} {\bibfnamefont {O.}~\bibnamefont {Kronenwerth}}, \bibinfo
  {author} {\bibfnamefont {C.}~\bibnamefont {Heyn}},\ and\ \bibinfo {author}
  {\bibfnamefont {D.}~\bibnamefont {Heitmann}},\ }\bibfield  {title} {\bibinfo
  {title} {Effect of the interface resistance on the extraordinary
  magnetoresistance of semiconductor/metal hybrid structures},\ }\href
  {https://doi.org/10.1023/A:1023246431624} {\bibfield  {journal} {\bibinfo
  {journal} {Journal of Superconductivity}\ }\textbf {\bibinfo {volume} {16}},\
  \bibinfo {pages} {195} (\bibinfo {year} {2003})}\BibitemShut {NoStop}%
\bibitem [{\citenamefont {Yang}\ \emph {et~al.}(2020)\citenamefont {Yang},
  \citenamefont {Liu}, \citenamefont {Fan},\ and\ \citenamefont
  {Zhang}}]{yang2020pms}%
  \BibitemOpen
  \bibfield  {author} {\bibinfo {author} {\bibfnamefont {M.}~\bibnamefont
  {Yang}}, \bibinfo {author} {\bibfnamefont {Y.}~\bibnamefont {Liu}}, \bibinfo
  {author} {\bibfnamefont {T.}~\bibnamefont {Fan}},\ and\ \bibinfo {author}
  {\bibfnamefont {D.}~\bibnamefont {Zhang}},\ }\bibfield  {title} {\bibinfo
  {title} {Metal-graphene interfaces in epitaxial and bulk systems: a review},\
  }\href {https://www.sciencedirect.com/science/article/pii/S0079642520300165}
  {\bibfield  {journal} {\bibinfo  {journal} {Progress in Materials Science}\
  }\textbf {\bibinfo {volume} {110}},\ \bibinfo {pages} {100652} (\bibinfo
  {year} {2020})}\BibitemShut {NoStop}%
\bibitem [{\citenamefont {Berdyugin}\ \emph {et~al.}(2019)\citenamefont
  {Berdyugin}, \citenamefont {Xu}, \citenamefont {Pellegrino}, \citenamefont
  {Krishna~Kumar}, \citenamefont {Principi}, \citenamefont {Torre},
  \citenamefont {Ben~Shalom}, \citenamefont {Taniguchi}, \citenamefont
  {Watanabe}, \citenamefont {Grigorieva}, \citenamefont {Polini}, \citenamefont
  {Geim},\ and\ \citenamefont {Bandurin}}]{berdyugin2019}%
  \BibitemOpen
  \bibfield  {author} {\bibinfo {author} {\bibfnamefont {A.~I.}\ \bibnamefont
  {Berdyugin}}, \bibinfo {author} {\bibfnamefont {S.~G.}\ \bibnamefont {Xu}},
  \bibinfo {author} {\bibfnamefont {F.~M.~D.}\ \bibnamefont {Pellegrino}},
  \bibinfo {author} {\bibfnamefont {R.}~\bibnamefont {Krishna~Kumar}}, \bibinfo
  {author} {\bibfnamefont {A.}~\bibnamefont {Principi}}, \bibinfo {author}
  {\bibfnamefont {I.}~\bibnamefont {Torre}}, \bibinfo {author} {\bibfnamefont
  {M.}~\bibnamefont {Ben~Shalom}}, \bibinfo {author} {\bibfnamefont
  {T.}~\bibnamefont {Taniguchi}}, \bibinfo {author} {\bibfnamefont
  {K.}~\bibnamefont {Watanabe}}, \bibinfo {author} {\bibfnamefont {I.~V.}\
  \bibnamefont {Grigorieva}}, \bibinfo {author} {\bibfnamefont
  {M.}~\bibnamefont {Polini}}, \bibinfo {author} {\bibfnamefont {A.~K.}\
  \bibnamefont {Geim}},\ and\ \bibinfo {author} {\bibfnamefont {D.~A.}\
  \bibnamefont {Bandurin}},\ }\bibfield  {title} {\bibinfo {title} {Measuring
  {{Hall}} viscosity of graphene’s electron fluid},\ }\href
  {https://doi.org/10.1126/science.aau0685} {\bibfield  {journal} {\bibinfo
  {journal} {Science}\ }\textbf {\bibinfo {volume} {364}},\ \bibinfo {pages}
  {162} (\bibinfo {year} {2019})}\BibitemShut {NoStop}%
\bibitem [{\citenamefont {Xin}\ \emph {et~al.}(2023)\citenamefont {Xin},
  \citenamefont {Lourembam}, \citenamefont {Kumaravadivel}, \citenamefont
  {Kazantsev}, \citenamefont {Wu}, \citenamefont {Mullan}, \citenamefont
  {Barrier}, \citenamefont {Geim}, \citenamefont {Grigorieva}, \citenamefont
  {Mishchenko} \emph {et~al.}}]{xin2023}%
  \BibitemOpen
  \bibfield  {author} {\bibinfo {author} {\bibfnamefont {N.}~\bibnamefont
  {Xin}}, \bibinfo {author} {\bibfnamefont {J.}~\bibnamefont {Lourembam}},
  \bibinfo {author} {\bibfnamefont {P.}~\bibnamefont {Kumaravadivel}}, \bibinfo
  {author} {\bibfnamefont {A.}~\bibnamefont {Kazantsev}}, \bibinfo {author}
  {\bibfnamefont {Z.}~\bibnamefont {Wu}}, \bibinfo {author} {\bibfnamefont
  {C.}~\bibnamefont {Mullan}}, \bibinfo {author} {\bibfnamefont
  {J.}~\bibnamefont {Barrier}}, \bibinfo {author} {\bibfnamefont {A.~A.}\
  \bibnamefont {Geim}}, \bibinfo {author} {\bibfnamefont {I.}~\bibnamefont
  {Grigorieva}}, \bibinfo {author} {\bibfnamefont {A.}~\bibnamefont
  {Mishchenko}}, \emph {et~al.},\ }\bibfield  {title} {\bibinfo {title} {Giant
  magnetoresistance of dirac plasma in high-mobility graphene},\ }\href@noop {}
  {\bibfield  {journal} {\bibinfo  {journal} {Nature}\ }\textbf {\bibinfo
  {volume} {616}},\ \bibinfo {pages} {270} (\bibinfo {year}
  {2023})}\BibitemShut {NoStop}%
\bibitem [{\citenamefont {Torre}\ \emph {et~al.}(2015)\citenamefont {Torre},
  \citenamefont {Tomadin}, \citenamefont {Geim},\ and\ \citenamefont
  {Polini}}]{torre2015}%
  \BibitemOpen
  \bibfield  {author} {\bibinfo {author} {\bibfnamefont {I.}~\bibnamefont
  {Torre}}, \bibinfo {author} {\bibfnamefont {A.}~\bibnamefont {Tomadin}},
  \bibinfo {author} {\bibfnamefont {A.~K.}\ \bibnamefont {Geim}},\ and\
  \bibinfo {author} {\bibfnamefont {M.}~\bibnamefont {Polini}},\ }\bibfield
  {title} {\bibinfo {title} {Nonlocal transport and the hydrodynamic shear
  viscosity in graphene},\ }\href {https://doi.org/10.1103/PhysRevB.92.165433}
  {\bibfield  {journal} {\bibinfo  {journal} {Phys. Rev. B}\ }\textbf {\bibinfo
  {volume} {92}},\ \bibinfo {pages} {165433} (\bibinfo {year}
  {2015})}\BibitemShut {NoStop}%
\bibitem [{\citenamefont {Nagaosa}\ \emph {et~al.}(2010)\citenamefont
  {Nagaosa}, \citenamefont {Sinova}, \citenamefont {Onoda}, \citenamefont
  {Macdonald},\ and\ \citenamefont {Ong}}]{nagaosa2010}%
  \BibitemOpen
  \bibfield  {author} {\bibinfo {author} {\bibfnamefont {N.}~\bibnamefont
  {Nagaosa}}, \bibinfo {author} {\bibfnamefont {J.}~\bibnamefont {Sinova}},
  \bibinfo {author} {\bibfnamefont {S.}~\bibnamefont {Onoda}}, \bibinfo
  {author} {\bibfnamefont {A.~H.}\ \bibnamefont {Macdonald}},\ and\ \bibinfo
  {author} {\bibfnamefont {N.~P.}\ \bibnamefont {Ong}},\ }\bibfield  {title}
  {\bibinfo {title} {Anomalous hall effect},\ }\href
  {https://doi.org/10.1103/RevModPhys.82.1539} {\bibfield  {journal} {\bibinfo
  {journal} {Reviews of Modern Physics}\ }\textbf {\bibinfo {volume} {82}},\
  \bibinfo {pages} {1539} (\bibinfo {year} {2010})}\BibitemShut {NoStop}%
\bibitem [{\citenamefont {von Klitzing}\ \emph {et~al.}(2020)\citenamefont {von
  Klitzing}, \citenamefont {Chakraborty}, \citenamefont {Kim}, \citenamefont
  {Madhavan}, \citenamefont {Dai}, \citenamefont {McIver}, \citenamefont
  {Tokura}, \citenamefont {Savary}, \citenamefont {Smirnova}, \citenamefont
  {Rey} \emph {et~al.}}]{von2020}%
  \BibitemOpen
  \bibfield  {author} {\bibinfo {author} {\bibfnamefont {K.}~\bibnamefont {von
  Klitzing}}, \bibinfo {author} {\bibfnamefont {T.}~\bibnamefont
  {Chakraborty}}, \bibinfo {author} {\bibfnamefont {P.}~\bibnamefont {Kim}},
  \bibinfo {author} {\bibfnamefont {V.}~\bibnamefont {Madhavan}}, \bibinfo
  {author} {\bibfnamefont {X.}~\bibnamefont {Dai}}, \bibinfo {author}
  {\bibfnamefont {J.}~\bibnamefont {McIver}}, \bibinfo {author} {\bibfnamefont
  {Y.}~\bibnamefont {Tokura}}, \bibinfo {author} {\bibfnamefont
  {L.}~\bibnamefont {Savary}}, \bibinfo {author} {\bibfnamefont
  {D.}~\bibnamefont {Smirnova}}, \bibinfo {author} {\bibfnamefont {A.~M.}\
  \bibnamefont {Rey}}, \emph {et~al.},\ }\bibfield  {title} {\bibinfo {title}
  {40 years of the quantum hall effect},\ }\href@noop {} {\bibfield  {journal}
  {\bibinfo  {journal} {Nature Reviews Physics}\ }\textbf {\bibinfo {volume}
  {2}},\ \bibinfo {pages} {397} (\bibinfo {year} {2020})}\BibitemShut {NoStop}%
\bibitem [{\citenamefont {Chopra}\ and\ \citenamefont
  {Hua}(2002)}]{chopra2002}%
  \BibitemOpen
  \bibfield  {author} {\bibinfo {author} {\bibfnamefont {H.~D.}\ \bibnamefont
  {Chopra}}\ and\ \bibinfo {author} {\bibfnamefont {S.~Z.}\ \bibnamefont
  {Hua}},\ }\bibfield  {title} {\bibinfo {title} {Ballistic magnetoresistance
  over 3000 in ni nanocontacts at room temperature},\ }\href
  {https://doi.org/10.1103/PhysRevB.66.020403} {\bibfield  {journal} {\bibinfo
  {journal} {Physical Review B}\ }\textbf {\bibinfo {volume} {66}},\ \bibinfo
  {pages} {020403} (\bibinfo {year} {2002})}\BibitemShut {NoStop}%
\bibitem [{\citenamefont {Caridad}\ \emph {et~al.}(2019)\citenamefont
  {Caridad}, \citenamefont {Power}, \citenamefont {Shylau}, \citenamefont
  {Gammelgaard}, \citenamefont {Jauho},\ and\ \citenamefont
  {Boggild}}]{caridad2019}%
  \BibitemOpen
  \bibfield  {author} {\bibinfo {author} {\bibfnamefont {J.~M.}\ \bibnamefont
  {Caridad}}, \bibinfo {author} {\bibfnamefont {S.~R.}\ \bibnamefont {Power}},
  \bibinfo {author} {\bibfnamefont {A.~A.}\ \bibnamefont {Shylau}}, \bibinfo
  {author} {\bibfnamefont {L.}~\bibnamefont {Gammelgaard}}, \bibinfo {author}
  {\bibfnamefont {A.}~\bibnamefont {Jauho}},\ and\ \bibinfo {author}
  {\bibfnamefont {P.}~\bibnamefont {Boggild}},\ }\bibfield  {title} {\bibinfo
  {title} {Gate electrostatics and quantum capacitance in ballistic graphene
  devices},\ }\href {https://doi.org/10.1103/PhysRevB.99.195408} {\bibfield
  {journal} {\bibinfo  {journal} {Physical Review B}\ }\textbf {\bibinfo
  {volume} {99}},\ \bibinfo {pages} {195408} (\bibinfo {year}
  {2019})}\BibitemShut {NoStop}%
\bibitem [{\citenamefont {Williams}\ \emph {et~al.}(2011)\citenamefont
  {Williams}, \citenamefont {Low}, \citenamefont {Lundstrom},\ and\
  \citenamefont {Marcus}}]{williams2011}%
  \BibitemOpen
  \bibfield  {author} {\bibinfo {author} {\bibfnamefont {J.~R.}\ \bibnamefont
  {Williams}}, \bibinfo {author} {\bibfnamefont {T.}~\bibnamefont {Low}},
  \bibinfo {author} {\bibfnamefont {M.~S.}\ \bibnamefont {Lundstrom}},\ and\
  \bibinfo {author} {\bibfnamefont {C.~M.}\ \bibnamefont {Marcus}},\ }\bibfield
   {title} {\bibinfo {title} {Gate-controlled guiding of electrons in
  graphene},\ }\href {https://doi.org/10.1038/nnano.2011.3} {\bibfield
  {journal} {\bibinfo  {journal} {Nat. Nanotech.}\ }\textbf {\bibinfo {volume}
  {6}},\ \bibinfo {pages} {222} (\bibinfo {year} {2011})}\BibitemShut {NoStop}%
\bibitem [{\citenamefont {Bhandari}\ \emph {et~al.}(2016)\citenamefont
  {Bhandari}, \citenamefont {Lee}, \citenamefont {Klales}, \citenamefont
  {Watanabe}, \citenamefont {Taniguchi}, \citenamefont {Heller}, \citenamefont
  {Kim},\ and\ \citenamefont {Westervelt}}]{bhandari2016}%
  \BibitemOpen
  \bibfield  {author} {\bibinfo {author} {\bibfnamefont {S.}~\bibnamefont
  {Bhandari}}, \bibinfo {author} {\bibfnamefont {G.~H.}\ \bibnamefont {Lee}},
  \bibinfo {author} {\bibfnamefont {A.}~\bibnamefont {Klales}}, \bibinfo
  {author} {\bibfnamefont {K.}~\bibnamefont {Watanabe}}, \bibinfo {author}
  {\bibfnamefont {T.}~\bibnamefont {Taniguchi}}, \bibinfo {author}
  {\bibfnamefont {E.}~\bibnamefont {Heller}}, \bibinfo {author} {\bibfnamefont
  {P.}~\bibnamefont {Kim}},\ and\ \bibinfo {author} {\bibfnamefont {R.~M.}\
  \bibnamefont {Westervelt}},\ }\bibfield  {title} {\bibinfo {title} {Imaging
  cyclotron orbits of electrons in graphene},\ }\href
  {https://doi.org/10.1021/acs.nanolett.5b04609} {\bibfield  {journal}
  {\bibinfo  {journal} {Nano Letters}\ }\textbf {\bibinfo {volume} {16}},\
  \bibinfo {pages} {1690} (\bibinfo {year} {2016})}\BibitemShut {NoStop}%
\bibitem [{\citenamefont {Taychatanapat}\ \emph {et~al.}(2013)\citenamefont
  {Taychatanapat}, \citenamefont {Watanabe}, \citenamefont {Taniguchi},\ and\
  \citenamefont {Jarillo-Herrero}}]{taychatanapat2013}%
  \BibitemOpen
  \bibfield  {author} {\bibinfo {author} {\bibfnamefont {T.}~\bibnamefont
  {Taychatanapat}}, \bibinfo {author} {\bibfnamefont {K.}~\bibnamefont
  {Watanabe}}, \bibinfo {author} {\bibfnamefont {T.}~\bibnamefont
  {Taniguchi}},\ and\ \bibinfo {author} {\bibfnamefont {P.}~\bibnamefont
  {Jarillo-Herrero}},\ }\bibfield  {title} {\bibinfo {title} {Electrically
  tunable transverse magnetic focusing in graphene},\ }\href@noop {} {\bibfield
   {journal} {\bibinfo  {journal} {Nature Physics}\ }\textbf {\bibinfo {volume}
  {9}},\ \bibinfo {pages} {225} (\bibinfo {year} {2013})}\BibitemShut {NoStop}%
\bibitem [{\citenamefont {Chen}\ \emph {et~al.}(2016)\citenamefont {Chen},
  \citenamefont {Han}, \citenamefont {Elahi}, \citenamefont {Habib},
  \citenamefont {Wang}, \citenamefont {Wen}, \citenamefont {Gao}, \citenamefont
  {Taniguchi}, \citenamefont {Watanabe}, \citenamefont {Hone}, \citenamefont
  {Ghosh},\ and\ \citenamefont {Dean}}]{chen2016science}%
  \BibitemOpen
  \bibfield  {author} {\bibinfo {author} {\bibfnamefont {S.~W.}\ \bibnamefont
  {Chen}}, \bibinfo {author} {\bibfnamefont {Z.}~\bibnamefont {Han}}, \bibinfo
  {author} {\bibfnamefont {M.~M.}\ \bibnamefont {Elahi}}, \bibinfo {author}
  {\bibfnamefont {K.~M.~M.}\ \bibnamefont {Habib}}, \bibinfo {author}
  {\bibfnamefont {L.}~\bibnamefont {Wang}}, \bibinfo {author} {\bibfnamefont
  {B.}~\bibnamefont {Wen}}, \bibinfo {author} {\bibfnamefont {Y.~D.}\
  \bibnamefont {Gao}}, \bibinfo {author} {\bibfnamefont {T.}~\bibnamefont
  {Taniguchi}}, \bibinfo {author} {\bibfnamefont {K.}~\bibnamefont {Watanabe}},
  \bibinfo {author} {\bibfnamefont {J.}~\bibnamefont {Hone}}, \bibinfo {author}
  {\bibfnamefont {A.~W.}\ \bibnamefont {Ghosh}},\ and\ \bibinfo {author}
  {\bibfnamefont {C.~R.}\ \bibnamefont {Dean}},\ }\bibfield  {title} {\bibinfo
  {title} {Electron optics with p-n junctions in ballistic graphene},\ }\href
  {https://doi.org/10.1126/science.aaf5481} {\bibfield  {journal} {\bibinfo
  {journal} {Science}\ }\textbf {\bibinfo {volume} {353}},\ \bibinfo {pages}
  {1522} (\bibinfo {year} {2016})}\BibitemShut {NoStop}%
\bibitem [{\citenamefont {Tao}\ \emph {et~al.}(2011)\citenamefont {Tao},
  \citenamefont {Jiao}, \citenamefont {Yazyev}, \citenamefont {Chen},
  \citenamefont {Feng}, \citenamefont {Zhang}, \citenamefont {Capaz},
  \citenamefont {Tour}, \citenamefont {Zettl}, \citenamefont {Louie},
  \citenamefont {Dai},\ and\ \citenamefont {Crommie}}]{tao2011}%
  \BibitemOpen
  \bibfield  {author} {\bibinfo {author} {\bibfnamefont {C.~G.}\ \bibnamefont
  {Tao}}, \bibinfo {author} {\bibfnamefont {L.~Y.}\ \bibnamefont {Jiao}},
  \bibinfo {author} {\bibfnamefont {O.~V.}\ \bibnamefont {Yazyev}}, \bibinfo
  {author} {\bibfnamefont {Y.~C.}\ \bibnamefont {Chen}}, \bibinfo {author}
  {\bibfnamefont {J.~J.}\ \bibnamefont {Feng}}, \bibinfo {author}
  {\bibfnamefont {X.~W.}\ \bibnamefont {Zhang}}, \bibinfo {author}
  {\bibfnamefont {R.~B.}\ \bibnamefont {Capaz}}, \bibinfo {author}
  {\bibfnamefont {J.~M.}\ \bibnamefont {Tour}}, \bibinfo {author}
  {\bibfnamefont {A.}~\bibnamefont {Zettl}}, \bibinfo {author} {\bibfnamefont
  {S.~G.}\ \bibnamefont {Louie}}, \bibinfo {author} {\bibfnamefont {H.~J.}\
  \bibnamefont {Dai}},\ and\ \bibinfo {author} {\bibfnamefont {M.~F.}\
  \bibnamefont {Crommie}},\ }\bibfield  {title} {\bibinfo {title} {Spatially
  resolving edge states of chiral graphene nanoribbons},\ }\href
  {https://doi.org/10.1038/NPHYS1991} {\bibfield  {journal} {\bibinfo
  {journal} {Nat. Phys.}\ }\textbf {\bibinfo {volume} {7}},\ \bibinfo {pages}
  {616} (\bibinfo {year} {2011})}\BibitemShut {NoStop}%
\bibitem [{\citenamefont {Bhandari}\ \emph {et~al.}(2020)\citenamefont
  {Bhandari}, \citenamefont {Lee}, \citenamefont {Watanabe}, \citenamefont
  {Taniguchi}, \citenamefont {Kim},\ and\ \citenamefont
  {Westervelt}}]{bhandari2020}%
  \BibitemOpen
  \bibfield  {author} {\bibinfo {author} {\bibfnamefont {S.}~\bibnamefont
  {Bhandari}}, \bibinfo {author} {\bibfnamefont {G.}~\bibnamefont {Lee}},
  \bibinfo {author} {\bibfnamefont {K.}~\bibnamefont {Watanabe}}, \bibinfo
  {author} {\bibfnamefont {T.}~\bibnamefont {Taniguchi}}, \bibinfo {author}
  {\bibfnamefont {P.}~\bibnamefont {Kim}},\ and\ \bibinfo {author}
  {\bibfnamefont {R.~M.}\ \bibnamefont {Westervelt}},\ }\bibfield  {title}
  {\bibinfo {title} {Imaging andreev reflection in graphene},\ }\href
  {https://doi.org/10.1021/acs.nanolett.0c00903} {\bibfield  {journal}
  {\bibinfo  {journal} {Nano Letters}\ }\textbf {\bibinfo {volume} {20}},\
  \bibinfo {pages} {4890} (\bibinfo {year} {2020})}\BibitemShut {NoStop}%
\bibitem [{\citenamefont {Mayorov}\ \emph {et~al.}(2011)\citenamefont
  {Mayorov}, \citenamefont {Gorbachev}, \citenamefont {Morozov}, \citenamefont
  {Britnell}, \citenamefont {Jalil}, \citenamefont {Ponomarenko}, \citenamefont
  {Blake}, \citenamefont {Novoselov}, \citenamefont {Watanabe}, \citenamefont
  {Taniguchi},\ and\ \citenamefont {Geim}}]{mayorov2011}%
  \BibitemOpen
  \bibfield  {author} {\bibinfo {author} {\bibfnamefont {A.~S.}\ \bibnamefont
  {Mayorov}}, \bibinfo {author} {\bibfnamefont {R.~V.}\ \bibnamefont
  {Gorbachev}}, \bibinfo {author} {\bibfnamefont {S.~V.}\ \bibnamefont
  {Morozov}}, \bibinfo {author} {\bibfnamefont {L.}~\bibnamefont {Britnell}},
  \bibinfo {author} {\bibfnamefont {R.}~\bibnamefont {Jalil}}, \bibinfo
  {author} {\bibfnamefont {L.~A.}\ \bibnamefont {Ponomarenko}}, \bibinfo
  {author} {\bibfnamefont {P.}~\bibnamefont {Blake}}, \bibinfo {author}
  {\bibfnamefont {K.~S.}\ \bibnamefont {Novoselov}}, \bibinfo {author}
  {\bibfnamefont {K.}~\bibnamefont {Watanabe}}, \bibinfo {author}
  {\bibfnamefont {T.}~\bibnamefont {Taniguchi}},\ and\ \bibinfo {author}
  {\bibfnamefont {A.~K.}\ \bibnamefont {Geim}},\ }\bibfield  {title} {\bibinfo
  {title} {Micrometer-{Scale} {Ballistic} {Transport} in {Encapsulated}
  {Graphene} at {Room} {Temperature}},\ }\href
  {https://doi.org/10.1021/nl200758b} {\bibfield  {journal} {\bibinfo
  {journal} {Nano Letters}\ }\textbf {\bibinfo {volume} {11}},\ \bibinfo
  {pages} {2396} (\bibinfo {year} {2011})}\BibitemShut {NoStop}%
\bibitem [{\citenamefont {Xu}\ \emph {et~al.}(2025)\citenamefont {Xu},
  \citenamefont {Palm}, \citenamefont {Huxter}, \citenamefont {Herb},
  \citenamefont {Abendroth}, \citenamefont {Bouzehouane}, \citenamefont
  {Boulle}, \citenamefont {Gabor}, \citenamefont {Urrestarazu-Larranaga},
  \citenamefont {Morales}, \citenamefont {Rhensius}, \citenamefont
  {Puebla-Hellmann},\ and\ \citenamefont {Degen}}]{xu2025}%
  \BibitemOpen
  \bibfield  {author} {\bibinfo {author} {\bibfnamefont {Z.}~\bibnamefont
  {Xu}}, \bibinfo {author} {\bibfnamefont {M.~L.}\ \bibnamefont {Palm}},
  \bibinfo {author} {\bibfnamefont {W.~S.}\ \bibnamefont {Huxter}}, \bibinfo
  {author} {\bibfnamefont {K.}~\bibnamefont {Herb}}, \bibinfo {author}
  {\bibfnamefont {J.~M.}\ \bibnamefont {Abendroth}}, \bibinfo {author}
  {\bibfnamefont {K.}~\bibnamefont {Bouzehouane}}, \bibinfo {author}
  {\bibfnamefont {O.}~\bibnamefont {Boulle}}, \bibinfo {author} {\bibfnamefont
  {M.}~\bibnamefont {Gabor}}, \bibinfo {author} {\bibfnamefont
  {J.}~\bibnamefont {Urrestarazu-Larranaga}}, \bibinfo {author} {\bibfnamefont
  {A.}~\bibnamefont {Morales}}, \bibinfo {author} {\bibfnamefont
  {J.}~\bibnamefont {Rhensius}}, \bibinfo {author} {\bibfnamefont {G.~F.}\
  \bibnamefont {Puebla-Hellmann}},\ and\ \bibinfo {author} {\bibfnamefont
  {C.~L.}\ \bibnamefont {Degen}},\ }\bibfield  {title} {\bibinfo {title}
  {Minimizing sensor-sample distances in scanning nitrogen-vacancy
  magnetometry},\ }\href {https://doi.org/10.1021/acsnano.4c18460} {\bibfield
  {journal} {\bibinfo  {journal} {ACS Nano}\ }\textbf {\bibinfo {volume}
  {19}},\ \bibinfo {pages} {8255} (\bibinfo {year} {2025})}\BibitemShut
  {NoStop}%
\bibitem [{\citenamefont {Butler}\ \emph {et~al.}(2013)\citenamefont {Butler},
  \citenamefont {Hollen}, \citenamefont {Cao}, \citenamefont {Cui},
  \citenamefont {Gupta}, \citenamefont {Gutierrez}, \citenamefont {Heinz},
  \citenamefont {Hong}, \citenamefont {Huang}, \citenamefont {Ismach},
  \citenamefont {Johnston-halperin}, \citenamefont {Kuno}, \citenamefont
  {Plashnitsa}, \citenamefont {Robinson}, \citenamefont {Ruoff}, \citenamefont
  {Salahuddin}, \citenamefont {Shan}, \citenamefont {Shi}, \citenamefont
  {Spencer}, \citenamefont {Terrones}, \citenamefont {Windl},\ and\
  \citenamefont {Goldberger}}]{butler2013}%
  \BibitemOpen
  \bibfield  {author} {\bibinfo {author} {\bibfnamefont {S.~Z.}\ \bibnamefont
  {Butler}}, \bibinfo {author} {\bibfnamefont {S.~M.}\ \bibnamefont {Hollen}},
  \bibinfo {author} {\bibfnamefont {L.~Y.}\ \bibnamefont {Cao}}, \bibinfo
  {author} {\bibfnamefont {Y.}~\bibnamefont {Cui}}, \bibinfo {author}
  {\bibfnamefont {J.~A.}\ \bibnamefont {Gupta}}, \bibinfo {author}
  {\bibfnamefont {H.~R.}\ \bibnamefont {Gutierrez}}, \bibinfo {author}
  {\bibfnamefont {T.~F.}\ \bibnamefont {Heinz}}, \bibinfo {author}
  {\bibfnamefont {S.~S.}\ \bibnamefont {Hong}}, \bibinfo {author}
  {\bibfnamefont {J.~X.}\ \bibnamefont {Huang}}, \bibinfo {author}
  {\bibfnamefont {A.~F.}\ \bibnamefont {Ismach}}, \bibinfo {author}
  {\bibfnamefont {E.}~\bibnamefont {Johnston-halperin}}, \bibinfo {author}
  {\bibfnamefont {M.}~\bibnamefont {Kuno}}, \bibinfo {author} {\bibfnamefont
  {V.~V.}\ \bibnamefont {Plashnitsa}}, \bibinfo {author} {\bibfnamefont
  {R.~D.}\ \bibnamefont {Robinson}}, \bibinfo {author} {\bibfnamefont {R.~S.}\
  \bibnamefont {Ruoff}}, \bibinfo {author} {\bibfnamefont {S.}~\bibnamefont
  {Salahuddin}}, \bibinfo {author} {\bibfnamefont {J.}~\bibnamefont {Shan}},
  \bibinfo {author} {\bibfnamefont {L.}~\bibnamefont {Shi}}, \bibinfo {author}
  {\bibfnamefont {M.~G.}\ \bibnamefont {Spencer}}, \bibinfo {author}
  {\bibfnamefont {M.}~\bibnamefont {Terrones}}, \bibinfo {author}
  {\bibfnamefont {W.}~\bibnamefont {Windl}},\ and\ \bibinfo {author}
  {\bibfnamefont {J.~E.}\ \bibnamefont {Goldberger}},\ }\bibfield  {title}
  {\bibinfo {title} {Progress challenges and opportunities in two-dimensional
  materials beyond graphene},\ }\href {https://doi.org/10.1021/nn400280c}
  {\bibfield  {journal} {\bibinfo  {journal} {ACS Nano}\ }\textbf {\bibinfo
  {volume} {7}},\ \bibinfo {pages} {2898} (\bibinfo {year} {2013})}\BibitemShut
  {NoStop}%
\bibitem [{\citenamefont {Nowack}\ \emph {et~al.}(2013)\citenamefont {Nowack},
  \citenamefont {Spanton}, \citenamefont {Baenninger}, \citenamefont {Konig},
  \citenamefont {Kirtley}, \citenamefont {Kalisky}, \citenamefont {Ames},
  \citenamefont {Leubner}, \citenamefont {Brune}, \citenamefont {Buhmann},
  \citenamefont {Molenkamp}, \citenamefont {Goldhaber-Gordon},\ and\
  \citenamefont {Moler}}]{nowack2013}%
  \BibitemOpen
  \bibfield  {author} {\bibinfo {author} {\bibfnamefont {K.~C.}\ \bibnamefont
  {Nowack}}, \bibinfo {author} {\bibfnamefont {E.~M.}\ \bibnamefont {Spanton}},
  \bibinfo {author} {\bibfnamefont {M.}~\bibnamefont {Baenninger}}, \bibinfo
  {author} {\bibfnamefont {M.}~\bibnamefont {Konig}}, \bibinfo {author}
  {\bibfnamefont {J.~R.}\ \bibnamefont {Kirtley}}, \bibinfo {author}
  {\bibfnamefont {B.}~\bibnamefont {Kalisky}}, \bibinfo {author} {\bibfnamefont
  {C.}~\bibnamefont {Ames}}, \bibinfo {author} {\bibfnamefont {P.}~\bibnamefont
  {Leubner}}, \bibinfo {author} {\bibfnamefont {C.}~\bibnamefont {Brune}},
  \bibinfo {author} {\bibfnamefont {H.}~\bibnamefont {Buhmann}}, \bibinfo
  {author} {\bibfnamefont {L.~W.}\ \bibnamefont {Molenkamp}}, \bibinfo {author}
  {\bibfnamefont {D.}~\bibnamefont {Goldhaber-Gordon}},\ and\ \bibinfo {author}
  {\bibfnamefont {K.~A.}\ \bibnamefont {Moler}},\ }\bibfield  {title} {\bibinfo
  {title} {Imaging currents in {{HgTe}} quantum wells in the quantum spin hall
  regime},\ }\href {https://doi.org/10.1038/nmat3682} {\bibfield  {journal}
  {\bibinfo  {journal} {Nat. Mater.}\ }\textbf {\bibinfo {volume} {12}},\
  \bibinfo {pages} {787} (\bibinfo {year} {2013})}\BibitemShut {NoStop}%
\bibitem [{\citenamefont {Jenkins}\ \emph {et~al.}(2022)\citenamefont
  {Jenkins}, \citenamefont {Baumann}, \citenamefont {Zhou}, \citenamefont
  {Meynell}, \citenamefont {Yang}, \citenamefont {Watanabe}, \citenamefont
  {Taniguchi}, \citenamefont {Lucas}, \citenamefont {Young},\ and\
  \citenamefont {Jayich}}]{jenkins2022}%
  \BibitemOpen
  \bibfield  {author} {\bibinfo {author} {\bibfnamefont {A.}~\bibnamefont
  {Jenkins}}, \bibinfo {author} {\bibfnamefont {S.}~\bibnamefont {Baumann}},
  \bibinfo {author} {\bibfnamefont {H.}~\bibnamefont {Zhou}}, \bibinfo {author}
  {\bibfnamefont {S.~A.}\ \bibnamefont {Meynell}}, \bibinfo {author}
  {\bibfnamefont {D.}~\bibnamefont {Yang}}, \bibinfo {author} {\bibfnamefont
  {K.}~\bibnamefont {Watanabe}}, \bibinfo {author} {\bibfnamefont
  {T.}~\bibnamefont {Taniguchi}}, \bibinfo {author} {\bibfnamefont
  {A.}~\bibnamefont {Lucas}}, \bibinfo {author} {\bibfnamefont {A.~F.}\
  \bibnamefont {Young}},\ and\ \bibinfo {author} {\bibfnamefont {A.~C.~B.}\
  \bibnamefont {Jayich}},\ }\bibfield  {title} {\bibinfo {title} {Imaging the
  breakdown of ohmic transport in graphene},\ }\href
  {https://doi.org/10.1103/PhysRevLett.129.087701} {\bibfield  {journal}
  {\bibinfo  {journal} {Physical Review Letters}\ }\textbf {\bibinfo {volume}
  {129}},\ \bibinfo {pages} {087701} (\bibinfo {year} {2022})}\BibitemShut
  {NoStop}%
\bibitem [{\citenamefont {Aharon-Steinberg}\ \emph {et~al.}(2022)\citenamefont
  {Aharon-Steinberg}, \citenamefont {V{\"o}lkl}, \citenamefont {Kaplan},
  \citenamefont {Pariari}, \citenamefont {Roy}, \citenamefont {Holder},
  \citenamefont {Wolf}, \citenamefont {Meltzer}, \citenamefont {Myasoedov},
  \citenamefont {Huber} \emph {et~al.}}]{aharon2022}%
  \BibitemOpen
  \bibfield  {author} {\bibinfo {author} {\bibfnamefont {A.}~\bibnamefont
  {Aharon-Steinberg}}, \bibinfo {author} {\bibfnamefont {T.}~\bibnamefont
  {V{\"o}lkl}}, \bibinfo {author} {\bibfnamefont {A.}~\bibnamefont {Kaplan}},
  \bibinfo {author} {\bibfnamefont {A.~K.}\ \bibnamefont {Pariari}}, \bibinfo
  {author} {\bibfnamefont {I.}~\bibnamefont {Roy}}, \bibinfo {author}
  {\bibfnamefont {T.}~\bibnamefont {Holder}}, \bibinfo {author} {\bibfnamefont
  {Y.}~\bibnamefont {Wolf}}, \bibinfo {author} {\bibfnamefont {A.~Y.}\
  \bibnamefont {Meltzer}}, \bibinfo {author} {\bibfnamefont {Y.}~\bibnamefont
  {Myasoedov}}, \bibinfo {author} {\bibfnamefont {M.~E.}\ \bibnamefont
  {Huber}}, \emph {et~al.},\ }\bibfield  {title} {\bibinfo {title} {Direct
  observation of vortices in an electron fluid},\ }\href@noop {} {\bibfield
  {journal} {\bibinfo  {journal} {Nature}\ }\textbf {\bibinfo {volume} {607}},\
  \bibinfo {pages} {74} (\bibinfo {year} {2022})}\BibitemShut {NoStop}%
\bibitem [{\citenamefont {Ferguson}\ \emph {et~al.}(2023)\citenamefont
  {Ferguson}, \citenamefont {Xiao}, \citenamefont {Richardella}, \citenamefont
  {Low}, \citenamefont {Samarth},\ and\ \citenamefont {Nowack}}]{ferguson2023}%
  \BibitemOpen
  \bibfield  {author} {\bibinfo {author} {\bibfnamefont {G.}~\bibnamefont
  {Ferguson}}, \bibinfo {author} {\bibfnamefont {R.}~\bibnamefont {Xiao}},
  \bibinfo {author} {\bibfnamefont {A.~R.}\ \bibnamefont {Richardella}},
  \bibinfo {author} {\bibfnamefont {D.}~\bibnamefont {Low}}, \bibinfo {author}
  {\bibfnamefont {N.}~\bibnamefont {Samarth}},\ and\ \bibinfo {author}
  {\bibfnamefont {K.~C.}\ \bibnamefont {Nowack}},\ }\bibfield  {title}
  {\bibinfo {title} {Direct visualization of electronic transport in a quantum
  anomalous hall insulator},\ }\href@noop {} {\bibfield  {journal} {\bibinfo
  {journal} {Nature Materials}\ }\textbf {\bibinfo {volume} {22}},\ \bibinfo
  {pages} {1100} (\bibinfo {year} {2023})}\BibitemShut {NoStop}%
\bibitem [{\citenamefont {Ferguson}\ \emph {et~al.}(2025)\citenamefont
  {Ferguson}, \citenamefont {Xiao}, \citenamefont {Richardella}, \citenamefont
  {Kaczmarek}, \citenamefont {Samarth},\ and\ \citenamefont
  {Nowack}}]{ferguson2025}%
  \BibitemOpen
  \bibfield  {author} {\bibinfo {author} {\bibfnamefont {G.~M.}\ \bibnamefont
  {Ferguson}}, \bibinfo {author} {\bibfnamefont {R.}~\bibnamefont {Xiao}},
  \bibinfo {author} {\bibfnamefont {A.~R.}\ \bibnamefont {Richardella}},
  \bibinfo {author} {\bibfnamefont {A.}~\bibnamefont {Kaczmarek}}, \bibinfo
  {author} {\bibfnamefont {N.}~\bibnamefont {Samarth}},\ and\ \bibinfo {author}
  {\bibfnamefont {K.~C.}\ \bibnamefont {Nowack}},\ }\bibfield  {title}
  {\bibinfo {title} {Imaging signatures of edge currents in a magnetic
  topological insulator},\ }\href {https://arxiv.org/abs/2501.11666} {\bibfield
   {journal} {\bibinfo  {journal} {arXiv:2501.11666}\ } (\bibinfo {year}
  {2025})}\BibitemShut {NoStop}%
\bibitem [{\citenamefont {Dockx}\ \emph {et~al.}(2025)\citenamefont {Dockx},
  \citenamefont {Buscema}, \citenamefont {Kumar}, \citenamefont {van Ree},
  \citenamefont {Mohtashami}, \citenamefont {van Dooren}, \citenamefont
  {Bulgarini}, \citenamefont {van Rijn}, \citenamefont {Osorio},\ and\
  \citenamefont {van~der Sar}}]{dockx2025}%
  \BibitemOpen
  \bibfield  {author} {\bibinfo {author} {\bibfnamefont {K.}~\bibnamefont
  {Dockx}}, \bibinfo {author} {\bibfnamefont {M.}~\bibnamefont {Buscema}},
  \bibinfo {author} {\bibfnamefont {S.}~\bibnamefont {Kumar}}, \bibinfo
  {author} {\bibfnamefont {T.}~\bibnamefont {van Ree}}, \bibinfo {author}
  {\bibfnamefont {A.}~\bibnamefont {Mohtashami}}, \bibinfo {author}
  {\bibfnamefont {L.}~\bibnamefont {van Dooren}}, \bibinfo {author}
  {\bibfnamefont {G.}~\bibnamefont {Bulgarini}}, \bibinfo {author}
  {\bibfnamefont {R.}~\bibnamefont {van Rijn}}, \bibinfo {author}
  {\bibfnamefont {C.~I.}\ \bibnamefont {Osorio}},\ and\ \bibinfo {author}
  {\bibfnamefont {T.}~\bibnamefont {van~der Sar}},\ }\bibfield  {title}
  {\bibinfo {title} {Imaging current flow and injection in scalable graphene
  devices through nv-magnetometry},\ }\href {https://arxiv.org/abs/2502.11076}
  {\bibfield  {journal} {\bibinfo  {journal} {arXiv:2502.11076}\ } (\bibinfo
  {year} {2025})}\BibitemShut {NoStop}%
\bibitem [{\citenamefont {Laturia}\ \emph {et~al.}(2018)\citenamefont
  {Laturia}, \citenamefont {Van~de Put},\ and\ \citenamefont
  {Vandenberghe}}]{laturia2018}%
  \BibitemOpen
  \bibfield  {author} {\bibinfo {author} {\bibfnamefont {A.}~\bibnamefont
  {Laturia}}, \bibinfo {author} {\bibfnamefont {M.~L.}\ \bibnamefont {Van~de
  Put}},\ and\ \bibinfo {author} {\bibfnamefont {W.~G.}\ \bibnamefont
  {Vandenberghe}},\ }\bibfield  {title} {\bibinfo {title} {Dielectric
  properties of hexagonal boron nitride and transition metal dichalcogenides:
  from monolayer to bulk},\ }\href@noop {} {\bibfield  {journal} {\bibinfo
  {journal} {npj 2D Materials and Applications}\ }\textbf {\bibinfo {volume}
  {2}},\ \bibinfo {pages} {6} (\bibinfo {year} {2018})}\BibitemShut {NoStop}%
\bibitem [{\citenamefont {Welter}(2022)}]{welter2022thesis}%
  \BibitemOpen
  \bibfield  {author} {\bibinfo {author} {\bibfnamefont {P.}~\bibnamefont
  {Welter}},\ }\bibfield  {title} {\bibinfo {title} {Microscopy of magnetic
  fields by scanning diamond magnetometry},\ }\href
  {https://doi.org/10.3929/ethz-b-000530838} {\bibfield  {journal} {\bibinfo
  {journal} {PhD Thesis, ETH Zurich, 2022}\ } (\bibinfo {year}
  {2022})}\BibitemShut {NoStop}%
\bibitem [{\citenamefont {Hewett}\ and\ \citenamefont
  {Kusmartsev}(2012)}]{hewett2012}%
  \BibitemOpen
  \bibfield  {author} {\bibinfo {author} {\bibfnamefont {T.~H.}\ \bibnamefont
  {Hewett}}\ and\ \bibinfo {author} {\bibfnamefont {F.~V.}\ \bibnamefont
  {Kusmartsev}},\ }\bibfield  {title} {\bibinfo {title} {Extraordinary
  magnetoresistance: sensing the future},\ }\href@noop {} {\bibfield  {journal}
  {\bibinfo  {journal} {Central European Journal of Physics}\ }\textbf
  {\bibinfo {volume} {10}},\ \bibinfo {pages} {602} (\bibinfo {year}
  {2012})}\BibitemShut {NoStop}%
\end{thebibliography}
